\documentclass{mn2e}
\def \th {\thinspace}
\def \degmark{^\circ}
\def \arcmin {\hbox{$^\prime$}}
\def \arcsec {\hbox{$^{\prime\prime}$}}

\def \arcmin {\hbox{$^\prime$}}
\def \arcsec {\hbox{$^{\prime\prime}$}}

\def\approxgt{\mathrel{\hbox{\rlap{\lower.55ex \hbox {$\sim$}}
  \kern-.3em \raise.4ex \hbox{$>$}}}}
\def\approxlt{\mathrel{\hbox{\rlap{\lower.55ex \hbox {$\sim$}}
  \kern-.3em \raise.4ex \hbox{$<$}}}}
\def \ref {\reference{}}
\def \sun {\hbox {$\odot$}}
\def \degmark{^\circ}

\usepackage{graphicx}
\usepackage{longtable}
\usepackage{subfigure}

\title[ULX in Normal Galaxies]             
         {Ultra-Luminous X-ray Source
Populations in Normal Galaxies: a Preliminary Survey with
Chandra}

\author[P. J. Humphrey et al.]    
      {P. J. Humphrey,$^1$
\newauthor      
      G. Fabbiano,$^2$ M. Elvis,$^2$
\newauthor
      M. J. Church$^{1,3}$ and M. Ba\l uci\'nska-Church$^{1,3}$\\
      $^1$University of Birmingham, School of Physics and Astronomy,
      Birmingham, B15 2TT, UK\\
      $^2$Harvard-Smithsonian Center for Astrophysics, 60 Garden
      Street, Cambridge MA 02138\\
      $^3$Astronomical Observatory, Jagiellonian University, ul. Orla 171,
      30-244 Cracow, Poland}

\date{Accepted 2003 May 15. Received 2002 June 7}

\begin{document}
\maketitle

\begin{abstract}
We present results of a {\it Chandra} survey of the
ultra-luminous X-ray sources (ULX) in 13 normal galaxies, in
which we combine source detection with X-ray flux measurement. 22
ULX were detected, i.e.  with $L_{\rm x}$ $>$ $\rm {1\times
10^{39}}$ erg s$^{-1}$ ($L_{10}$), and 39 other sources were
detected with $L_{\rm x}$ $>$ $\rm {5\times 10^{38}}$ erg
s$^{-1}$ ($L_5$).  We also use radial intensity profiles to
remove extended sources from the sample.  The majority of sources
are not extended, which for a typical distance constrains the
emission region size to less than 50 pc. X-ray colour-colour
diagrams and spectral fitting results were examined for
indicators of the ULX nature.  In the case of the brighter
sources, spectral fitting generally requires two-component
models. In only a few cases do colour-colour diagrams or spectral
fitting provide evidence of black hole nature.  
We find no evidence of a correlation with stellar mass, however there is
a strong correlation with star formation as indicated by
the 60 $\mu$m flux as found in previous studies.
\end{abstract}

\begin{keywords}
                accretion: accretion discs --
                binaries: close --
                black hole physics --
                X-rays: galaxies --
                X-rays: stars
\end{keywords}

\section{Introduction}

Ultra-luminous X-ray sources (ULX) are intriguing, apparently
point-like sources in external galaxies which are normally
distinguished from bright central objects.  Bright objects may be
divided into those that exceed the Eddington limit $L_{\rm Edd}$
for a 1.4 M$_{\sun}$ neutron star of $\sim$2--4${\rm \times
10^{38}}$ erg s$^{-1}$, and those with luminosity greater than
$10^{39}$ erg s$^{-1}$, which is the normal definition of an
ULX. Observations of bright spiral galaxies with {\it Einstein}
revealed substantial numbers of very bright X-ray sources
external to the nuclear regions, i.e. 36 super-Eddington sources
of which 16 were ULX (Fabbiano 1989). A significant fraction of
the X-ray emission from galaxies has long been known to originate
from X-ray binaries (e.g. Fabbiano 1989; Fabbiano, Kim \&
Trinchieri 1994, Fabbiano et al. 2001, Blanton et al. 2001).  In
a survey of nearby galaxies with {\it Rosat}, Roberts \& Warwick
(2000) found 28 ULX outside the nuclei. Similarly, Colbert \&
Mushotzky (1999) investigated extra-nuclear ($>$2~arcmin offset)
X-ray sources in 39 nearby galaxies finding 13 ULX.  They
suggested these were accreting black hole systems of
10$^2$--10$^4$ M$_{\sun}$.  The spectra of 7 ULX in nearby
galaxies studied using {\it ASCA} by Makishima et al. (2000) were
well-fitted by a multi-colour disc blackbody model suggesting
black hole nature.  {\it Chandra} observations of galaxies have
revealed a large number of previously unknown ULX, the
luminosities of which extend from $\rm {1-10\times 10^{39}}$ erg
s$^{-1}$ (Fabbiano, Zezas \& Murray 2001; Blanton et al. 2001),
suggesting that they may be quite common.  With the advent of
{\it Chandra}, it becomes possible to study ULX in more detail.

\tabcolsep 1.5mm
\begin{table*}
\begin{center}			
\begin{minipage}{126mm}
\caption{Galaxy sample and {\it Chandra} observation log.  ACIS
exposures are shown, and the minimum resolvable size scale
(equivalent to 0.5 arcsec) using our best distances for the
sources (Table 2). We also show the line-of-sight column
densities within our galaxy for the centroid of each external
galaxy.  (Dickey \& Lockman 1990).}
\label{galtable}
\begin{tabular}{llllllllrr}
\hline\noalign{\smallskip}
Galaxy & Type & $T$ & $B_0^{\rm T}$ &$D_{25}$ & $N_{\rm H}^{\rm GAL}$&ObsID & Date & Exp & Resln \\ 
& & & &$^\prime$ & 10$^{20}$ cm$^{-2}$& &dd/mm/yr &ks &pc\\
\hline\noalign{\smallskip}
NGC\th 4636 &E0-1 & -4.8 & $10.$ & 6.0 & 1.8 & 323 &  26/01/00& 52 & 36 \\
NGC\th 1132 &E & -4.8  &$13.$ & 2.5& 5.2 & 801& 10/12/99& 13 & 170 \\
NGC\th 4697 &E6 & -4.7 &  $10.$ & 7.2 & 2.2 &  784 & 15/01/00& 39 & 28 \\
NGC\th 1399 &E1 & -4.5 & $10.$ & 6.9 & 1.3 & 319 & 18/01/00& 56 & 50 \\
NGC\th 1291 &S0/a & 0.1 & $9.3$ & 9.8& 2.1 & 2059& 07/11/00& 23 &30 \\
NGC\th 2681 &S0/a & 0.4 & $11.$ & 3.6& 2.4 & 2060& 30/01/01& 77 & 42 \\
NGC\th 253  &Sc   & 5.1 & $7.1$ & 28 & 1.4 & 969 & 16/12/99& 14 & 7.5 \\
NGC\th 3184 &Scd & 5.9 & $10.$ & 7.4& 1.1& 804& 08/01/00& 41 & 17 \\
NGC\th 4631 &Sd & 6.5 & $8.6$ & 15& 1.3 & 797& 16/04/00& 59 & 8.5 \\
IC\th 5332 &Sd & 6.8 & $11.$ & 7.8& 1.4 & 2066 & 02/05/01 & 52 & 9.7 \\
IC\th 2574 &Sm & 8.9 & $10.$ & 13& 2.4 & 792& 07/01/00& 8.6 & 8.8 \\
NGC\th 1569 &Im & 9.6 & $9.4$ & 3.6& 2.2 & 782& 11/04/00& 92 & 4.1 \\
IZW\th 18 & \ldots &\ldots &\ldots & 0.30 & 1.9 & 805 & 08/02/00& 31 & 31 \\
\hline\noalign{\smallskip}
\end{tabular}
\end{minipage}
\end{center}
\end{table*}					

The discovery of ULX naturally led to the proposal that these may
be a single type of object. Explanations of this type have
involved firstly, an intermediate mass black hole
binary. Makishima et al. (2000) required black hole masses of
between $\sim$3 and $\sim$80 M$_{\sun}$ for the Eddington limit
not to be exceeded. In general, intermediate mass black hole
models involve masses of $10^2$ - $10^4$ M$_{\sun}$, i.e. more
massive than Galactic BHB such as Cyg\th X-1, and substantially
less massive than AGN. This possibility has been invoked to
explain the most luminous, variable ULX in M82 (Kaaret et
al. 2001). 
However, King et al. (2001) discussed the difficulties of forming these, and proposed 
an alternative model involving mild beaming and a link with Galactic micro-quasars.
It has been known that ULX occur preferentially in regions of star formation
(Zezas et al. 1999; Roberts \& Warwick 2000; Fabbiano et al. 2001), and the model 
is consistent with the expected association of high mass X-ray binaries (HMXB) with
young stellar populations. The numbers of ULX found in the Antennae galaxies 
(Fabbiano et al. 2001) supports the connection with recent massive star formation.  
ULX were also detected in elliptical galaxies which do not contain
HMXB, e.g. Sarazin et al. (2001), which led King (2002) to extend the model 
by proposing two types of ULX: persistent
sources predominating in galaxies with young stellar populations, and
micro-quasars with bright, prolonged outbursts occurring in elliptical galaxies.

However, there is also the possibility that ULX do not consist of
a single or even two types of object.  Objects generating X-ray
luminosities larger than the Eddington limit for a neutron star
include not only stellar mass black holes (M $<$ 100 M$_{\sun}$)
but very young supernova remnants (Roberts \& Warwick, 2000; King
et al., 2001).  Some pre-{\it Chandra} ULX may have been
unresolved ``super-bubbles'' of shock-heated HII in the ISM
(e.g. Stewart \& Walter 2000), typically with diameters
$\sim$200--1000~pc. With the 0.3 arcsec (on-axis) resolution of
{\it Chandra} (van Speybroeck et al. 1997; van Speybroeck 1999)
these objects would appear extended in nearby galaxies, for
example, 0.3 arcsec corresponds to a size of 27-40 pc at the
distance of the Antennae galaxies (20-30 Mpc) (Fabbiano et
al. 2001). Young, compact supernova remnants, which can reach
luminosities of a few $\times 10^{40}$ erg~s$^{-1}$ (Immler \&
Lewin 2002) would not be resolved even with {\it Chandra}, except
in the nearest members of the Local Group.  Similarly an
unresolved cluster of sub-Eddington low mass or high mass XRB may
appear ultra-luminous. This could be in a globular cluster such
as the two sources in the Galactic globular cluster M15, White \&
Angelini 2001). Since the diameter of a globular cluster is
typically 10 pc, these objects also cannot be resolved, even with
{\em Chandra}.  Large amplitude variability would be a strong
pointer that the source consists of a single object or a group
of a small number of objects only (Roberts \& Warwick, 2000).

In this paper, we examine systematically {\it Chandra} ACIS
observations of 13 normal galaxies to detect all super-Eddington
sources and so determine whether the prevalence of ULX depends
upon galaxy morphology.  One aim of the work was to examine the
spectra of the brighter sources and search for any spectral
property that may reveal the nature of a ULX source.

\section{The Galaxy Sample}

Thirteen galaxies were selected from the available {\it Chandra}
observations in the public archive in the summer of 2001, that
were classified by the {\it Chandra} X-ray Centre ({\it CXC}) as
Normal Galaxies.  These galaxies were selected on the basis that
the observations were longer than $\sim$10 ksec, and covered a
range of types.  While this is not a statistically complete
sample, the seven spiral, four elliptical and two irregular
galaxies allow us to investigate whether there is any correlation
of ULX occurrence with morphology. The galaxies chosen are listed
in in morphological order from E to Irr in Table 1, with details
of the observations.  Morphological types and face-on,
absorption-corrected $B$-band magnitude, $B_0^{\rm T}$ are taken
from the Third Reference Catalogue of Bright Galaxies (RC3; de
Vaucouleurs et al. 1991).  The morphological parameter $T$, where
-5 is Elliptical, 0 is S0a and 10 is Irregular, is taken from the
Lyon-Meudon Extragalactic Database {\it LEDA} (Paturel et
al. 1997), based on RC3.

{\em Chandra} point-source detections have already been reported
for NGC\th 1399 (Angelini, Loewenstein \& Mushotzky 2001), NGC\th
4697 (Sarazin, Irwin \& Bregman 2000, 2001) and the central
region of NGC\th 1291 (Irwin, Sarazin \& Bregman 2001). However,
since spectral studies of individual objects were not in general
made, these sources were included here.

\tabcolsep 3.0 mm
\begin{table}                       
\begin{center}
\begin{minipage}{80mm}
\caption{Preferred values of distances of each galaxy in Mpc\label{distances}}
\begin{tabular}{llrrr}
\hline\noalign{\smallskip}
&Galaxy&distance&technique&reference\\
\hline\noalign{\smallskip}
&NGC\th 4636 & $14.7\pm 0.9$&SBF &1\\  
&NGC\th 1132 &$69 \pm 5$    &$D_{\rm N}$-$\sigma$ &2\\
&NGC\th 4697 &$11.7\pm 0.8$  &SBF &1\\
&NGC\th 1399 &$20 \pm 1$    &SBF &1\\
&NGC\th 1291 &$12 \pm 3$    &$D_{\rm N}$-$\sigma$ &3 \\
&NGC\th 2681 &$17 \pm 3$    &SBF &1\\
&NGC\th 253  &$3.1\pm 0.7$   &Cepheid &4  \\
&NGC\th 3184 &$7.2\pm 1.7$   &T-F &4 \\
&NGC\th 4631 &$3.5\pm 0.3$   &T-F &4 \\
&IC\th 5332  &$4.0\pm 1.0$   &T-F &5 \\
&IC\th 2574  &$3.6\pm 0.3$   &Cepheid &4\\
&NGC\th 1569 &$1.7$          &BS&4\\
&IZW\th 18   &$12\pm $2    &$\rm V_{\rm vir}$&6 \\
\hline\noalign{\smallskip}
\end{tabular}

The techniques shown are: SBF: {\it I}-band surface brightness fluctuation;
$D_{\rm N}$-$\sigma$: isophotal diameter velocity dispersion; T-F: Tully-Fisher; 
BS: brightest stars in galaxies; $\rm V_{\rm vir}$: redshift distances with
Virgocentric correction. 

References: 
(1) Tonry et al. 2001;
(2) Djorgovski \& Davis 1987;
(3) Prugniel \& Simien 1996; 
(4) Shapley et al. 2001;
(5) Bottinelli 1985;
(6) \"Ostlin 2000; 
\end{minipage}
\end{center}
\end{table}                                       

Based on a literature search, we obtained preferred distances for
the 13 galaxies as shown in Table 2.  In the case of 5 galaxies,
we use the distances given by Shapley, Fabbiano \& Eskridge
(2001). For the other galaxies, we adopt, in order of preference
the values obtained from Cepheid variables; from $I$-band
surface-brightness fluctuations; from the Tully-Fisher relation;
or from the isophotal diameter - velocity dispersion ($D_{\rm
N}$--$\sigma$) relation.  Otherwise, we adopt estimates based on
the brightest star in the galaxy, or use the redshift corrected
for Virgocentric flow from {\it LEDA}. Distance errors are also
shown, reflecting the uncertainty in the most reliable technique
available for any galaxy.  Where possible, systematic and
statistical errors are combined.  If no errors were quoted in the
literature we adopt, if possible, typical values for the given
distance method as listed in Jacoby et al. (1992). For the
redshift-distances we adopt errors determined by Shapley et
al. (2001) of $\sim$30\% (20\% in the case of NGC\th 1132).

\section{X-ray Source Detection}

Data analysis was performed using the {\sc ciao} 2.1.2 software,
{\sc xanadu} and {\sc ftools} 5.0.  To remove periods of high
background, lightcurves were accumulated from source-free regions
of the active chips and intervals having factor of two increases
in count rate removed.  This led to significant data loss only
for IZW\th 18 and NGC\th 1291, in which $\sim$25\% and $\sim$35\%
of data were removed, respectively. Net exposure times are shown
in Table 1.

Searches for point sources were made using the {\sc ciao} {\sc
wavdetect} algorithm, without regard to whether detections were
at the centres of galaxies or not. However, as discussed below,
only a few detections coincided with galaxy centres.  The
spurious source detection probability was set at $10^{-6}$ per
pixel, corresponding to 0.014 spurious detections per square
arcmin.
Exposure maps created with {\sc mkexpmap} at 1.7~keV were used
with {\sc wavdetect} to minimize spurious detections at the
detector chip boundaries and these aspects were checked by manual
inspection of the images.  In two galaxies, IC\th 5332 and NGC\th
1569, no sources were detected exceeding $\rm {5 \times 10^{38}}$
erg~s$^{-1}$.
Detections spanning 2, 4, 8 or 16 pixels were
flagged, as these imply extended sources. Radial profiles
(Sect. 3.1) confirmed possible extended sources as those
significantly broader than the PSF. For a typical off-axis PSF of
1 arcsec width, a typical distance of 10 Mpc implies a size of 50 pc,
so that to appear point-like in {\it Chandra} a source must have
a size no larger than this. Only 10 sources
eventually proved to have an extended component.  

We selected the point sources lying within the $B$-band 25th
magnitude isophotes derived from RC3, which are thus likely to be
associated with the galaxy. For these, spectra, lightcurves and
instrument response functions were generated. Background data
were extracted from annuli centred upon the source region and
lying entirely within the same CCD node. The annuli were chosen
to cover an area at least 8 times larger than the source
extraction region and containing at least 20 photons (although
such regions frequently exceeded this limit). To prevent
contamination of the background spectra by photons from bright
point sources, all photons detected within a region six times
larger than the 1-$\sigma$ encircled-energy ellipse of every
source, and centred on that source were removed prior to
background accumulation.  Such a large masking region was adopted
since with a 3-$\sigma$ region, a few per cent of a source's
intensity will remain.

The total count from a given detection varied between $\sim$100 and 
1000 (Table 4). Thus to plot light curves in the normal way to achieve Poisson 
errors of 10\%, say, would allow a very small number of bins, and searching
for variability by $\chi^2$ testing would not be sensible in most cases.
We thus applied a Kolmogorov-Smirnov test to the arrival times
of photons and the numbers of photons that arrive by a particular time
are compared with those expected for constant intensity.
This test strictly requires unbinned data, and so we used the primitive binning
of the ACIS data of 3.24 s, although binning does not bias the test
provided there are many primitive bins. This revealed only one source
in galaxy NGC\th 253 which displayed variability at the 99\%
significance level, this source having a luminosity of $\rm {2.5
\times 10^{38}}$ erg s$^{-1}$, located at $\alpha$ = 00$^h$ 47$^m$ 30.$^s$9,
$\delta$ = -25$\degmark$ 18\arcmin $\,$ 26\arcsec (2000). Other sources exhibited no evidence for
variability, however the Kolmogorov-Smirnov test does not provide upper limits
for non-detections. We carried out simple $\chi^2$ testing which indicated
that other sources were not variable at a confidence of 
at least 90\%.
We adopt the procedure (below) of
considering only sources brighter than $\rm {5 \times 10^{38}}$
erg s$^{-1}$, so that no bright sources detected (shown in Tables
3 -- 5) displayed significant variability. Thus, none of the
sources in the sample examined provided evidence in this way that
the object was not a superposition of sources.

Radial brightness profiles were derived for each point source
detected, which allowed extended sources to be clearly identified
by comparing the profiles with the PSF derived for each position
(see Sect. 3.1). Ten sources found using {\sc wavdetect} were
rejected as being clearly extended, and 2 more sources found to
have point source and extended components are discussed in
Sect. 3.1.

Spectral fitting (Sect. 3.3) provided best-fit models from which
fluxes were derived.  Then, using the preferred distances of
Table 2, luminosities were calculated in the band 0.3--7.0 keV.
In the case of the weaker sources, which do not permit sensible
fitting to discriminate between models, an absorbed power law was
fitted with a fixed power law photon index of 2.0 (typical of fit
results in other cases) and column density fixed at the
line-of-sight Galactic value (Dickey \& Lockman 1990; see Table
1).  After removal of extended sources, in this band, 258 sources
were found with $L_{\rm x} >\rm {1\times 10^{38}}$ erg s$^{-1}$
($L_1$), of which 158 ($N_{2}$) had $L_{\rm x} > \rm {2\times
10^{38}}$ erg s$^{-1}$ ($L_2$), 61 ($N_{5}$) had $L_{\rm x} >\rm
{5\times 10^{38}}$ ergs$^{-1}$ ($L_5$), and 22 ($N_{10}$) had
$L_{\rm x}$ $>$ $\rm {1\times 10^{39}}$ erg s$^{-1}$ ($L_{10}$).

In Table 3 we show the number of sources detected in each galaxy,
$N_{2}$, $N_{5}$ and $N_{10}$. $N_{2}$ varied from 2 in NGC\th
1132 to 72 in NGC\th 1399. The numbers detected have error values
calculated using the errors in luminosity from spectral fitting
combined with the distance uncertainties.  In Fig. 1, Digital Sky
Survey images of each galaxy are shown with 25${\rm ^{th}}$
magnitude ($D_{25}$) {\it B}-band isophotes from RC3
superimposed. All sources with luminosity $>$ $L_{10}$ are shown,
together with sources lying between $L_{5}$ and $L_{10}$.

To estimate how many background objects may have been detected as
ULX, the Log\th N($>$S) - Log\th S relations in the {\it Chandra}
Deep Field South (Giacconi et al. 2001) in the energy bands
0.3--2.0 keV and 2.0--7.0 keV were used to obtain the expected
number of background objects within each galaxy (i.e. within the
25th magnitude isophote).  Data in these bands were combined into
a single plot for the total band, approximately, by using the
spectral form of Giacconi et al. (2001) to relate the total flux
in the 0.3--7.0 keV band to the fluxes in the sub-bands.  It was
assumed that soft sources were also detected in the hard band,
and so the correction was made using $N$ values from the hard
band and correcting $S$ values to the flux in the total band. The
number of background objects detected within each galaxy is shown
in Table 3. For the 13 galaxies, 17 spurious sources are expected
with apparent luminosities $>$ $\rm {5\times
10^{38}}$~erg~s$^{-1}$, and 8 with $L_{\rm x}$ $>$ $1\times
10^{39}$~erg~s$^{-1}$. Hence 62$\pm$7\% of the ULX sources are
true members of their host galaxies (Table~3).  To estimate the
errors of the numbers of sources in each luminosity-band, the
errors in distance and flux must be combined, bearing in mind
that the distance errors do not affect each source
independently. A simple Monte-Carlo method was adopted with 10000
simulations for each galaxy. Each simulation consisted first of
taking a random distance to the galaxy, normally distributed
about the measured value, with a standard deviation equal to the
1-$\sigma$ distance uncertainties in Table 2. From this distance,
appropriate flux limits for each luminosity-band were computed.
A similar procedure was adopted taking a random flux for each
detection distributed about its measured value, and the number of
sources exceeding each flux limit were counted. We adopted the
standard deviation of these numbers, accumulated over all the
simulations, as the error estimate.

We show the ULX detected in Fig. 1 (as diagonal crosses) in each
galaxy, and other super-Eddington sources ($L_{\rm x}$ $>$ $L_5$)
are shown separately.  In Table 4 (upper panel), individual ULX
in each galaxy are listed, and given a source name such as NGC\th
4636 PSX-1, the ``P'' indicating pointlike nature. For each
source, the total number of counts (`count') contained in the
spectrum (discussed in Sect. 3.3), the source count-rate, the
luminosity (L$_{\rm x}$) in the band 0.3--7.0 keV derived from
spectral fitting using the best distance value, with 90\%
confidence errors, the offset ($\Delta R$) of the source from the
nominal galaxy centre.  Source names following the ${\it
Chandra}$-naming convention are given. In the upper panel we
include sources
\renewcommand{\thesubfigure}{\thefigure.x}
\makeatletter
        \renewcommand{\@thesubfigure}{}
        \renewcommand{\p@subfigure}{}
\makeatother
\renewcommand{\subfigtopskip}{0pt}
\renewcommand{\subfigcapskip}{0pt}
\renewcommand{\subfigbottomskip}{10pt}
\begin{figure*}                                         
\centering
\subfigure[\hbox{NGC 4636 \hspace{43pt} E0--1}]{
\includegraphics[scale=0.22]{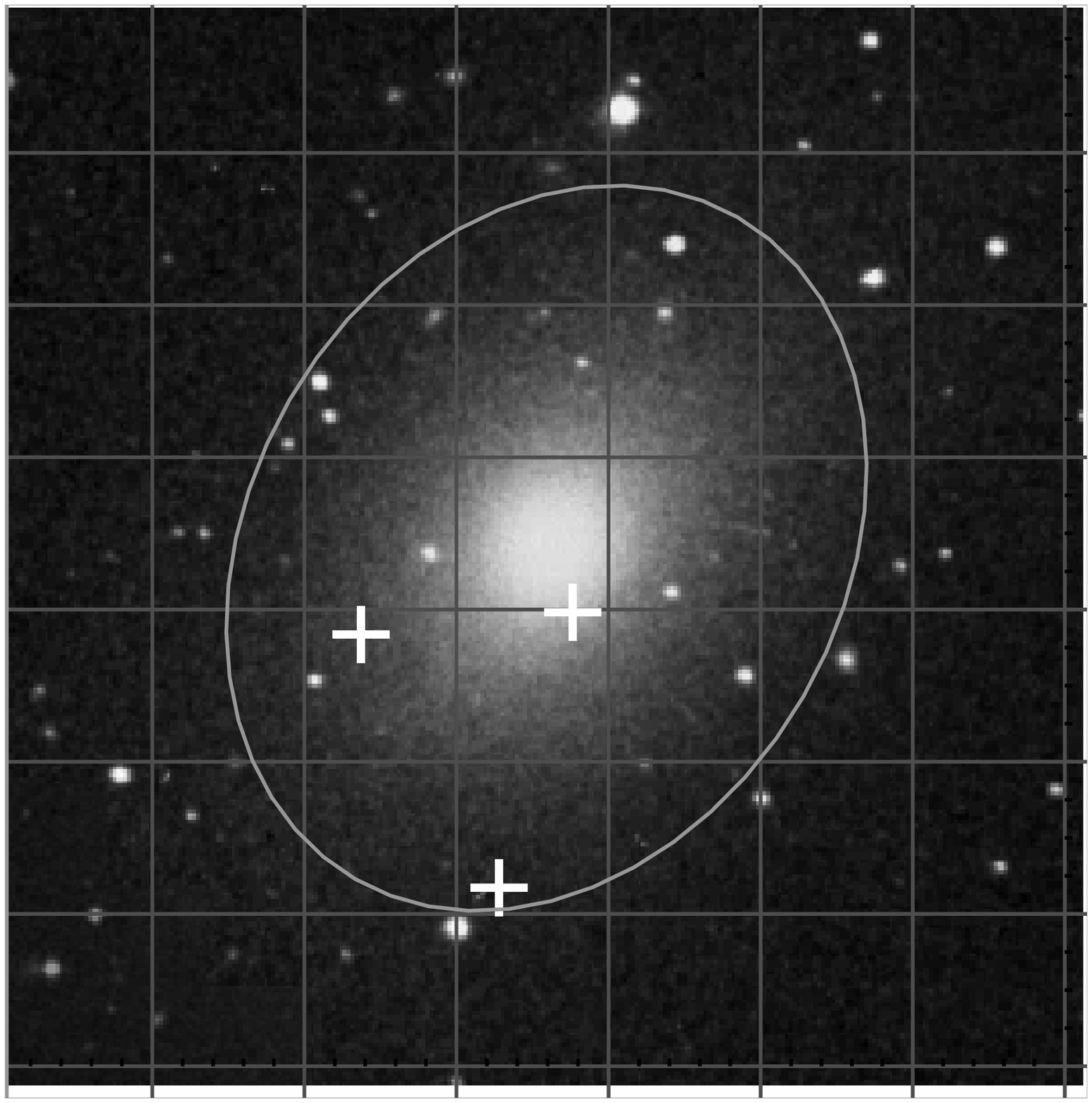}}
\subfigure[\hbox{NGC 1132 \hspace{50pt} E}]{\includegraphics[scale=0.22]{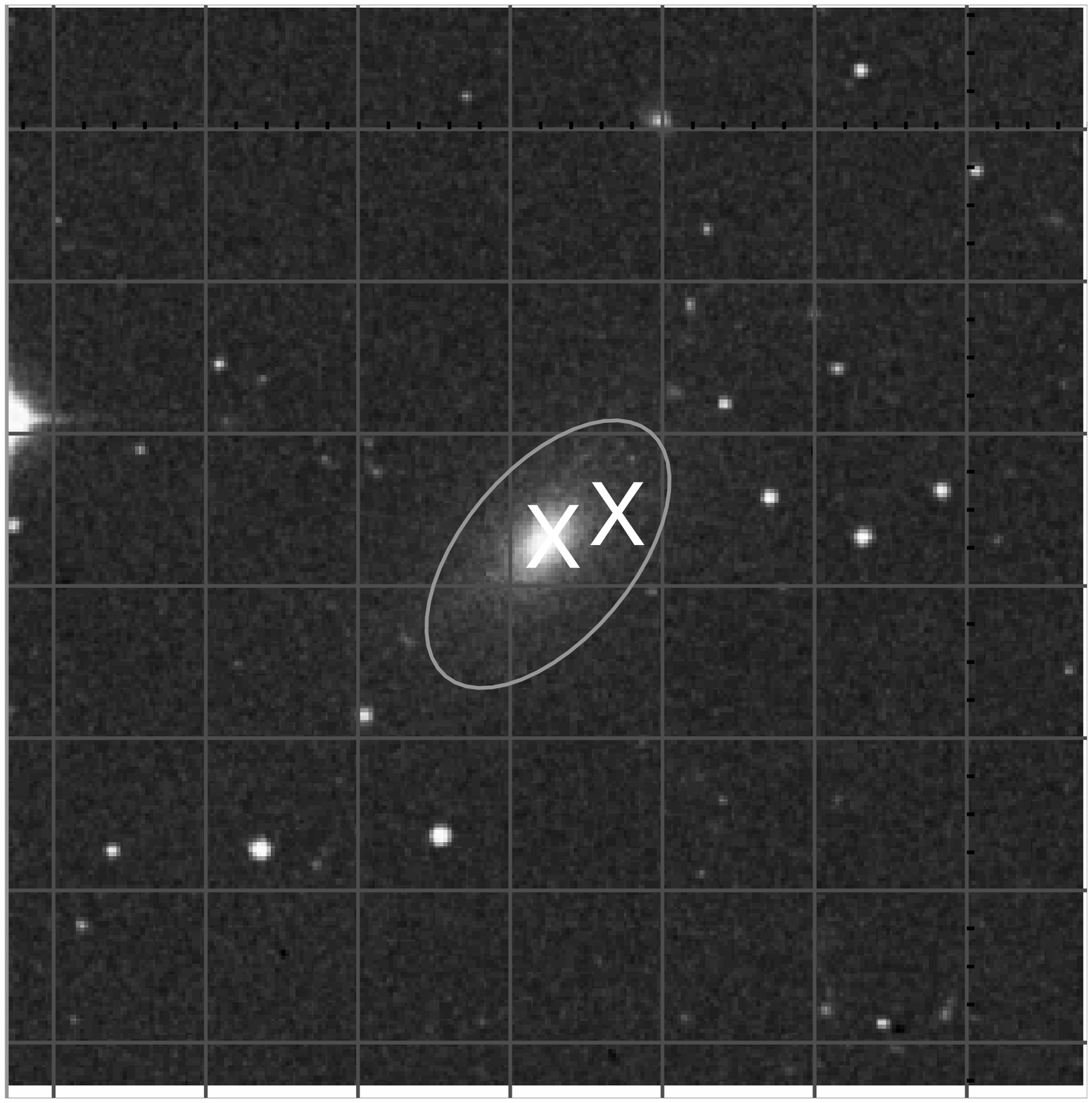}}
\subfigure[\hbox{NGC 4697 \hspace{46pt} E6}]{\includegraphics[scale=0.22]{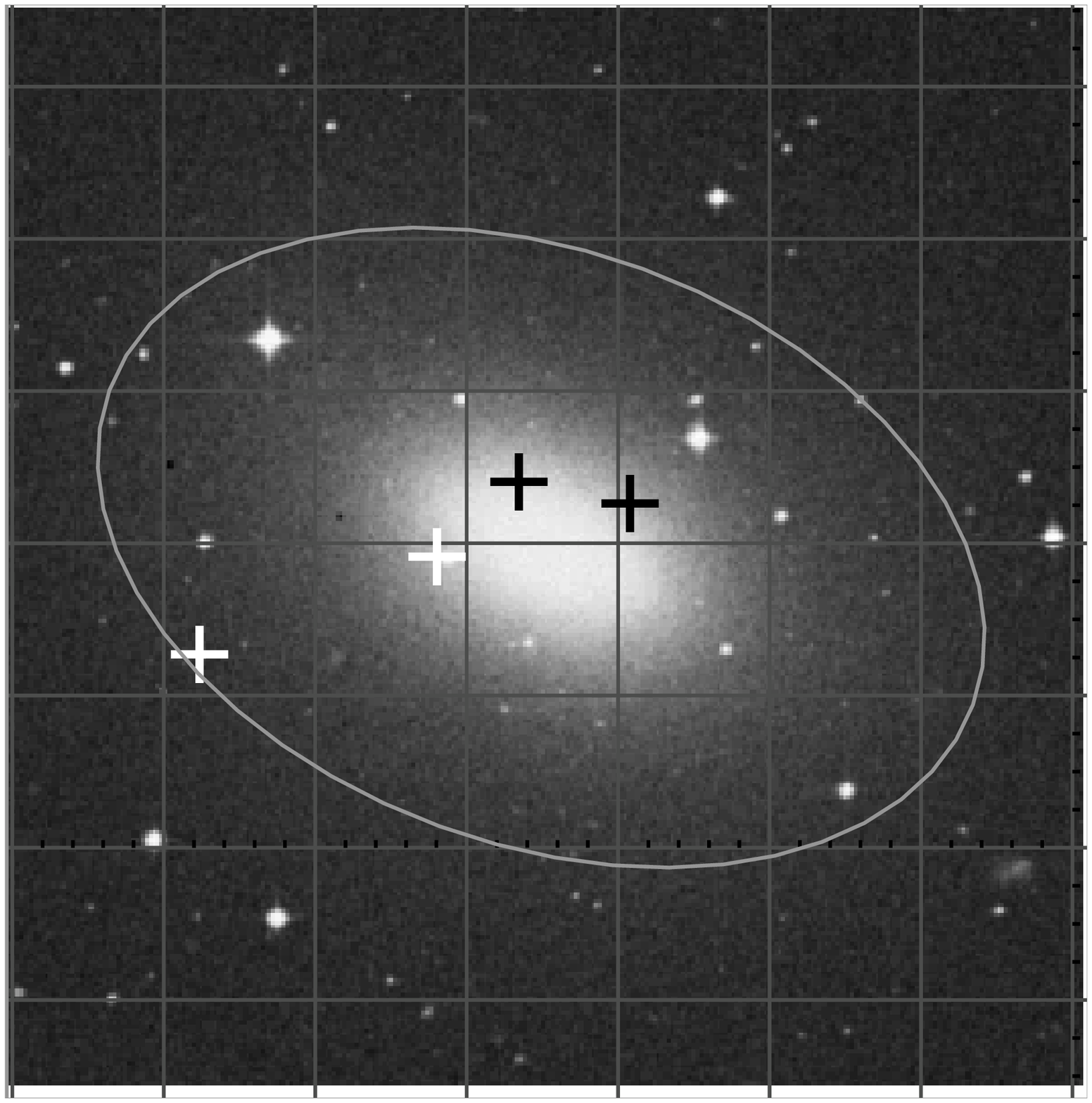}}
\subfigure[\hbox{NGC 1399 \hspace{46pt} E1}]{\includegraphics[scale=0.22]{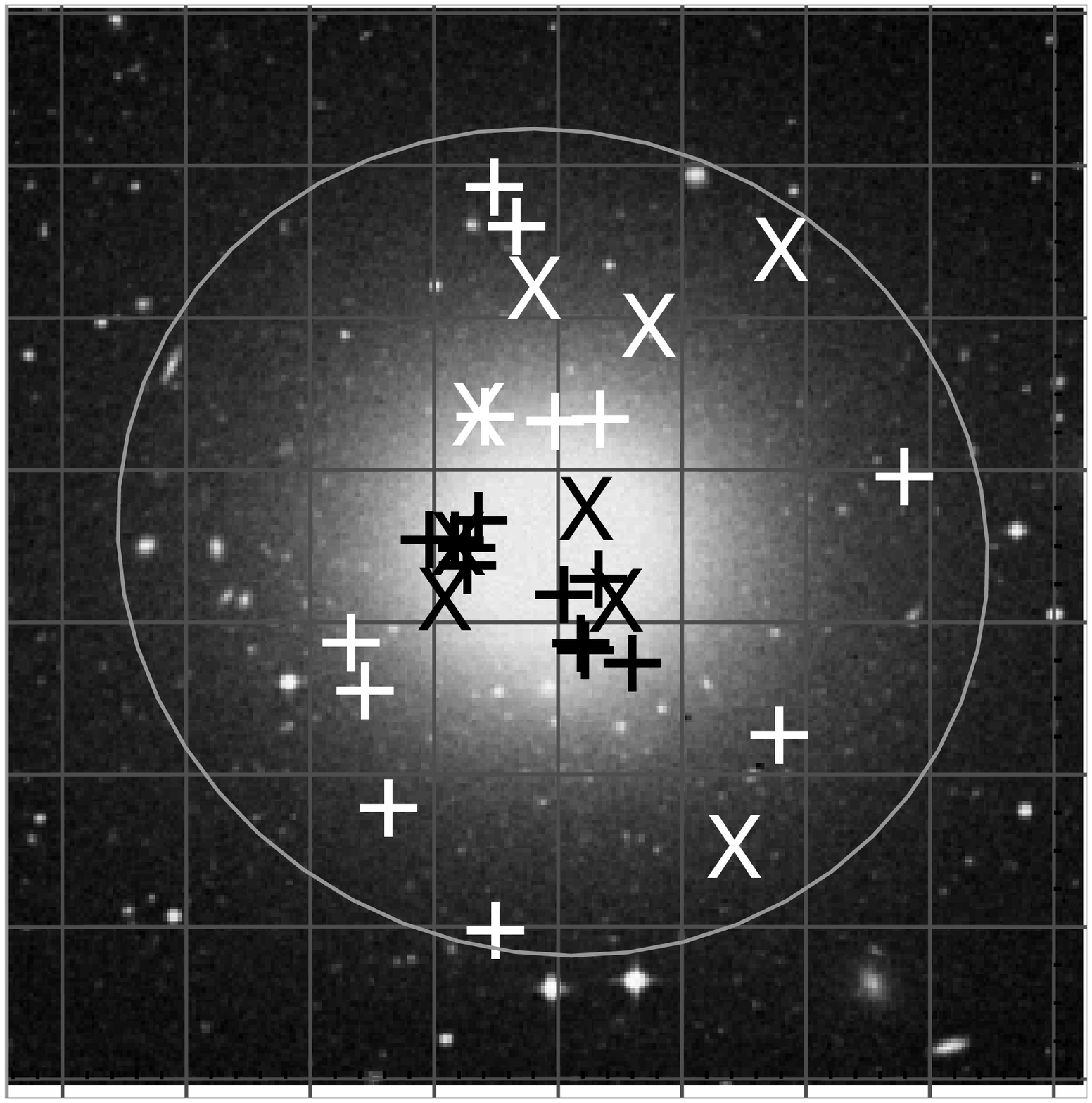}}
\subfigure[\hbox{NGC 1291 \hspace{43pt} S0/a}]{\includegraphics[scale=0.22]{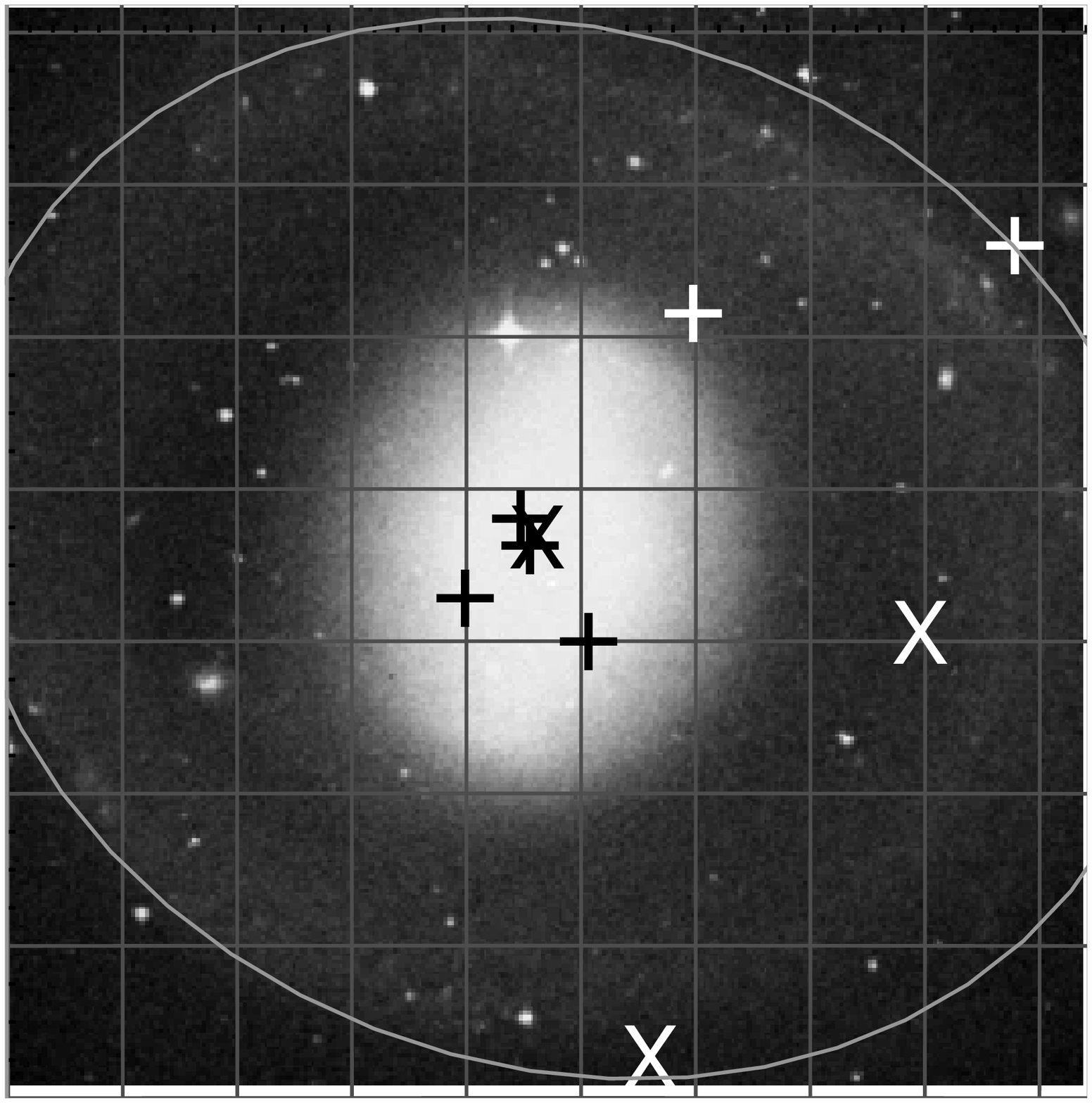}}
\subfigure[\hbox{NGC 2681 \hspace{43pt} S0/a}]{\includegraphics[scale=0.22]{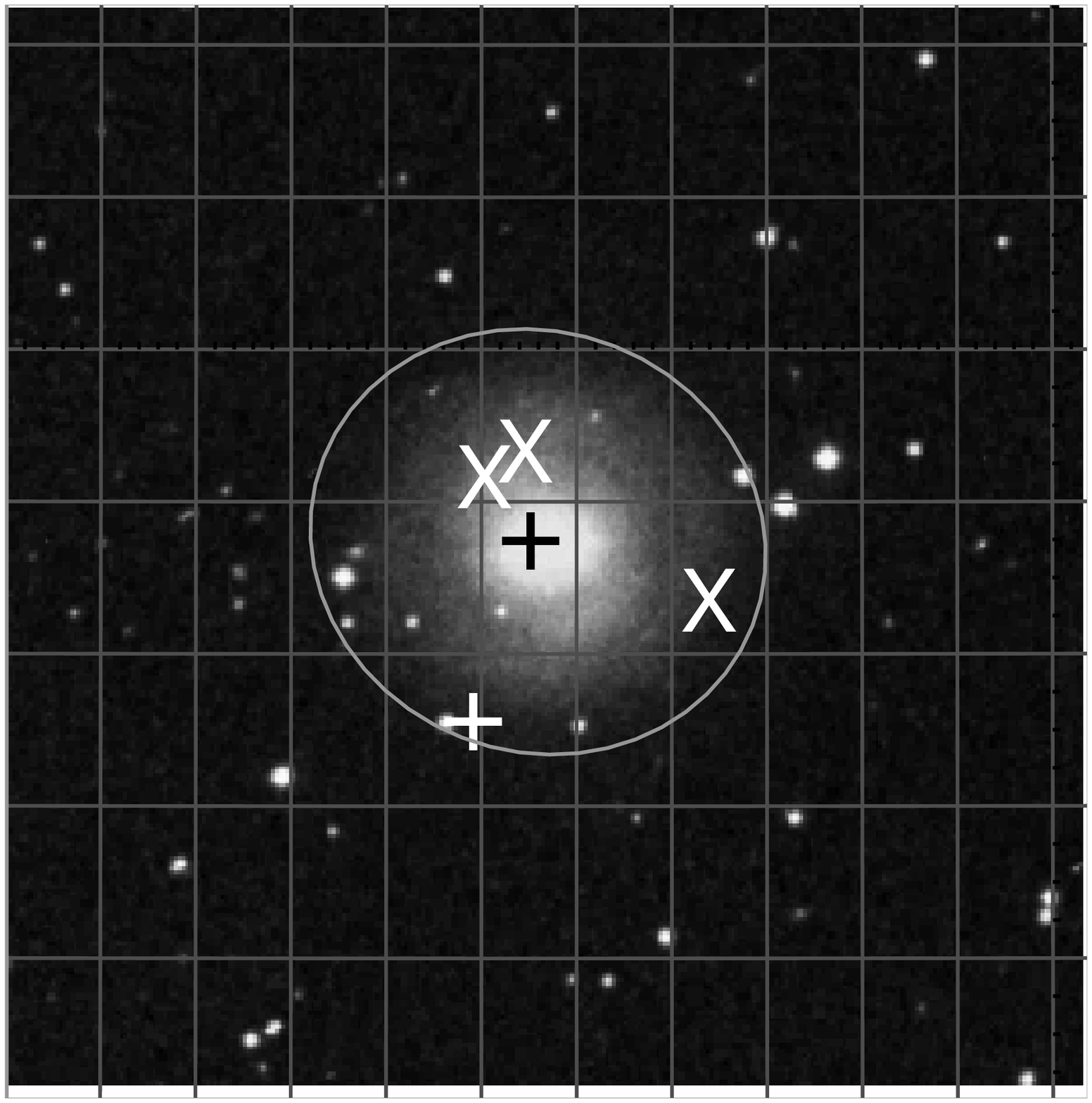}}
\subfigure[\hbox{NGC 253 \hspace{49pt} Sc}]{\includegraphics[scale=0.22]{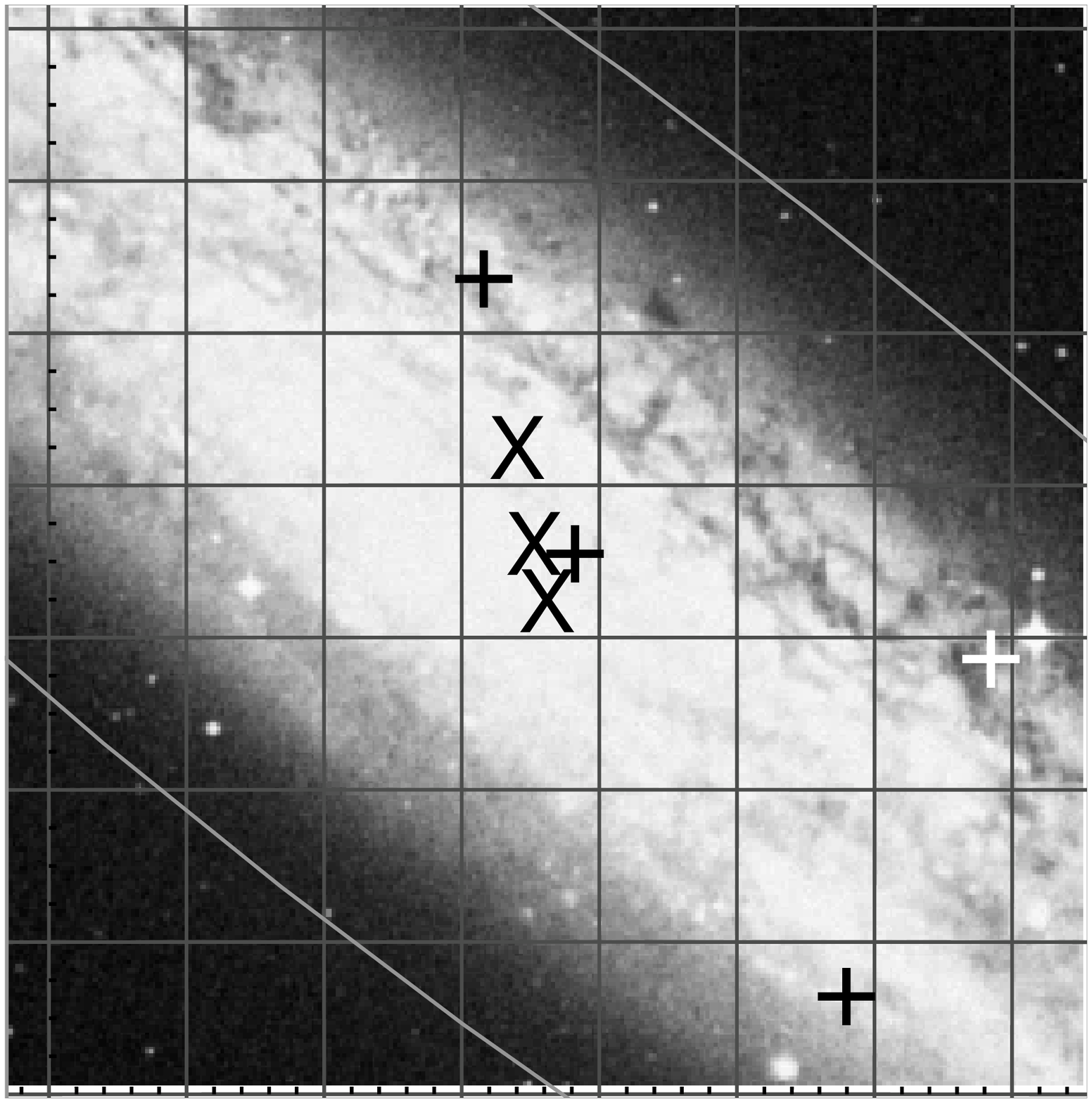}}
\subfigure[\hbox{NGC 3184 \hspace{43pt} Scd}]{\includegraphics[scale=0.22]{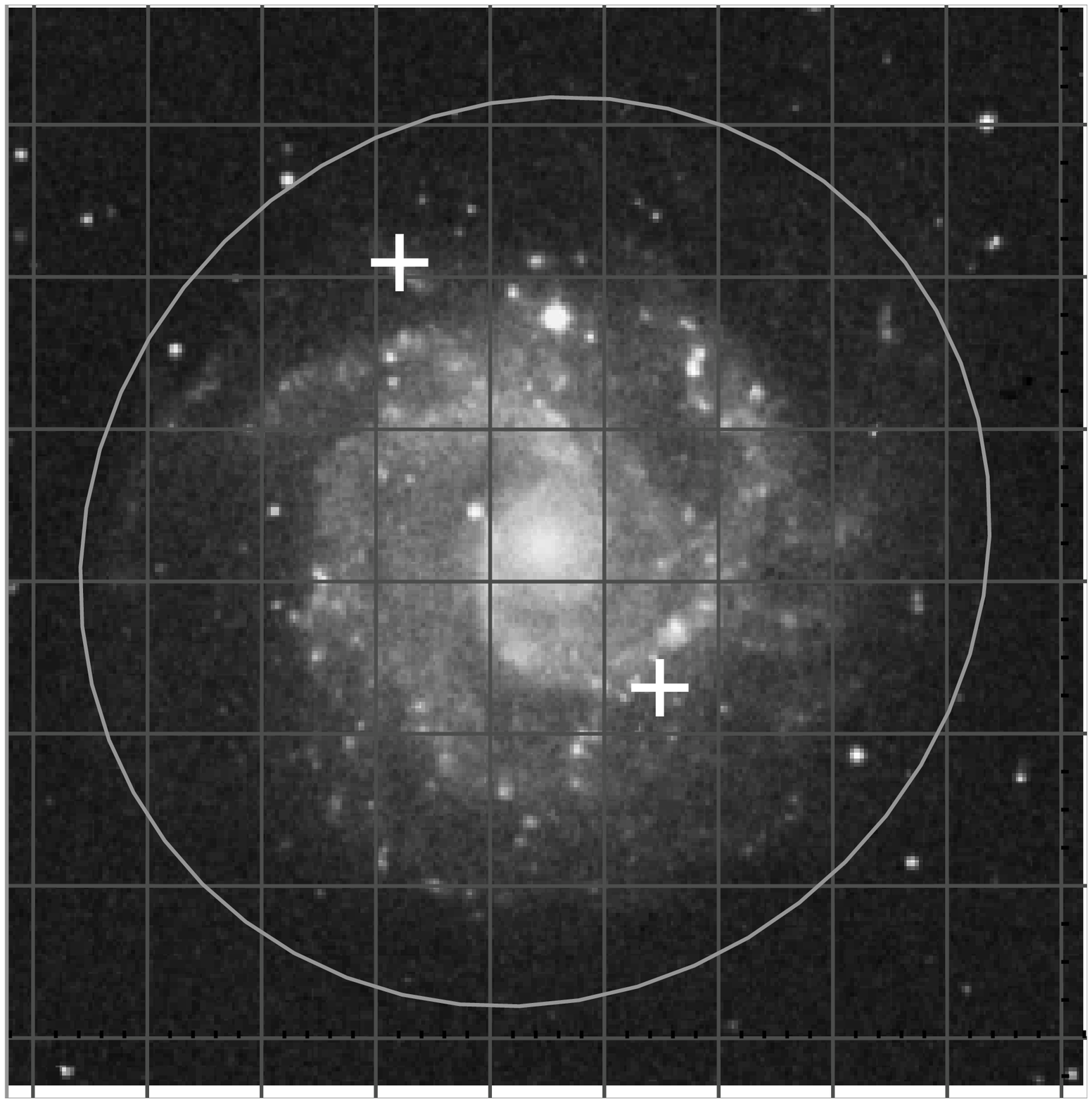}}
\subfigure[\hbox{NGC 4631 \hspace{46pt} Sd}]{\includegraphics[scale=0.22]{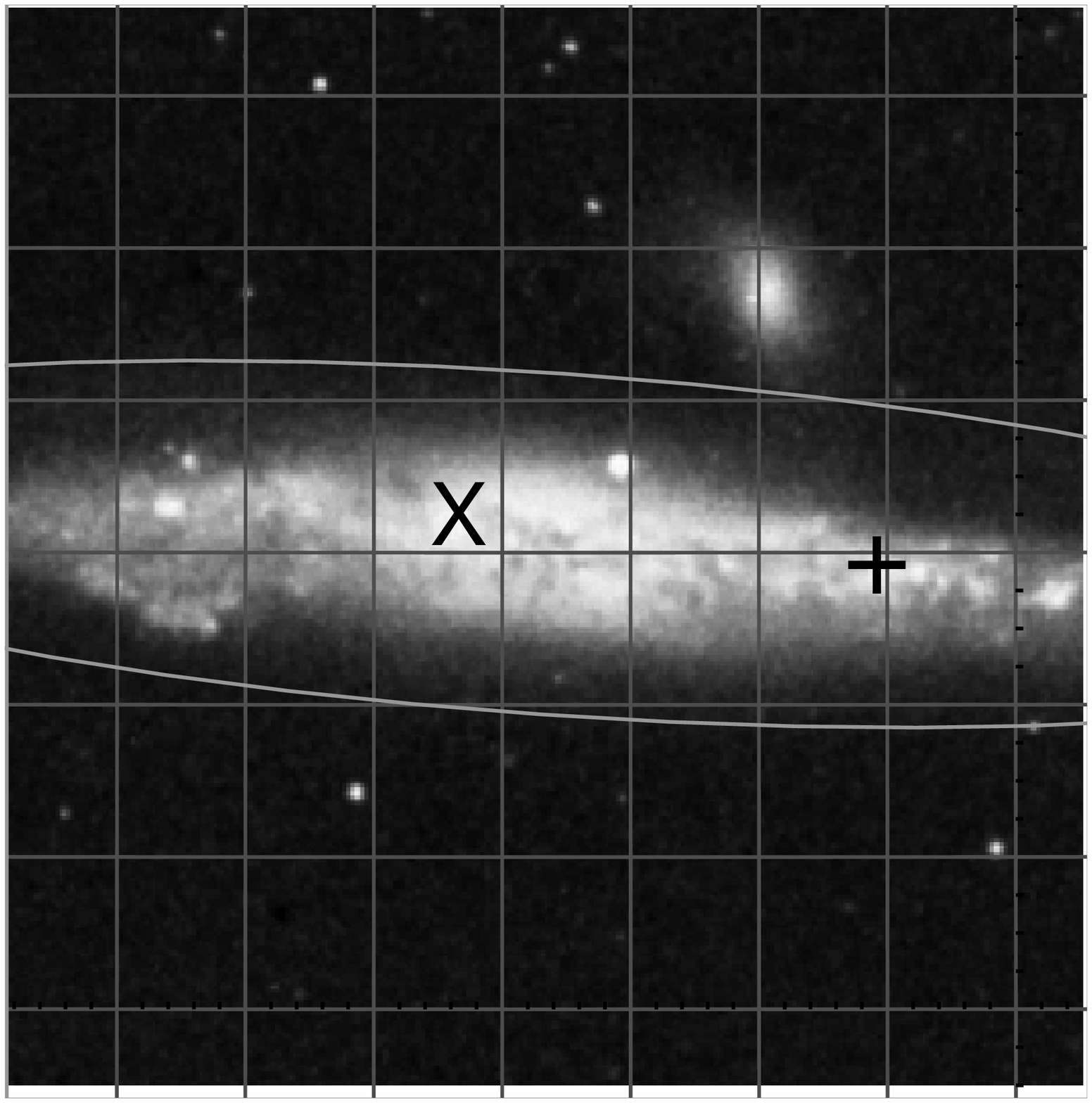}}
\subfigure[\hbox{IC 5332 \hspace{49pt} Sd}]{\includegraphics[scale=0.22]{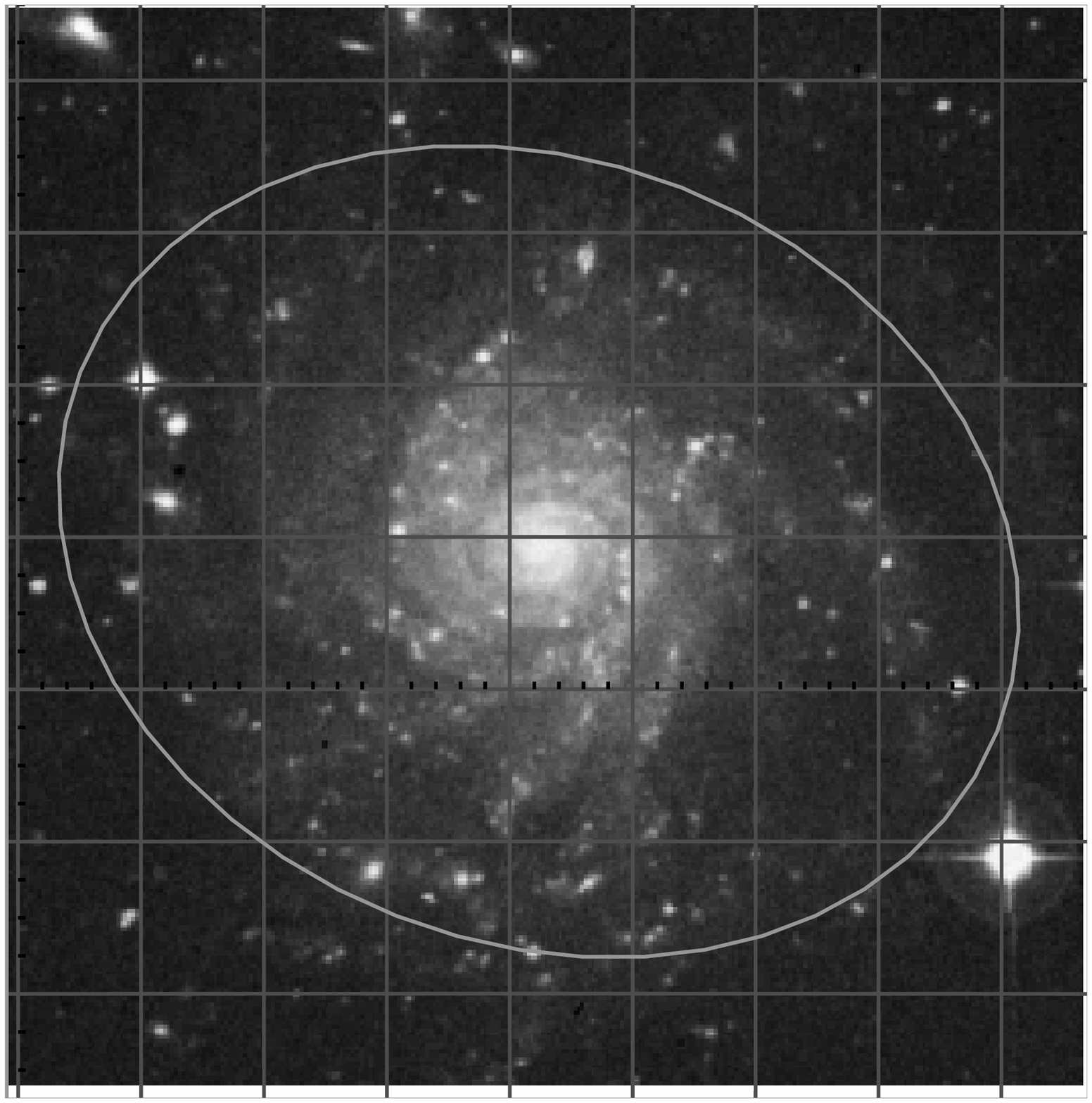}}
\subfigure[\hbox{IC 2574 \hspace{49pt} Sm}]{\includegraphics[scale=0.22]{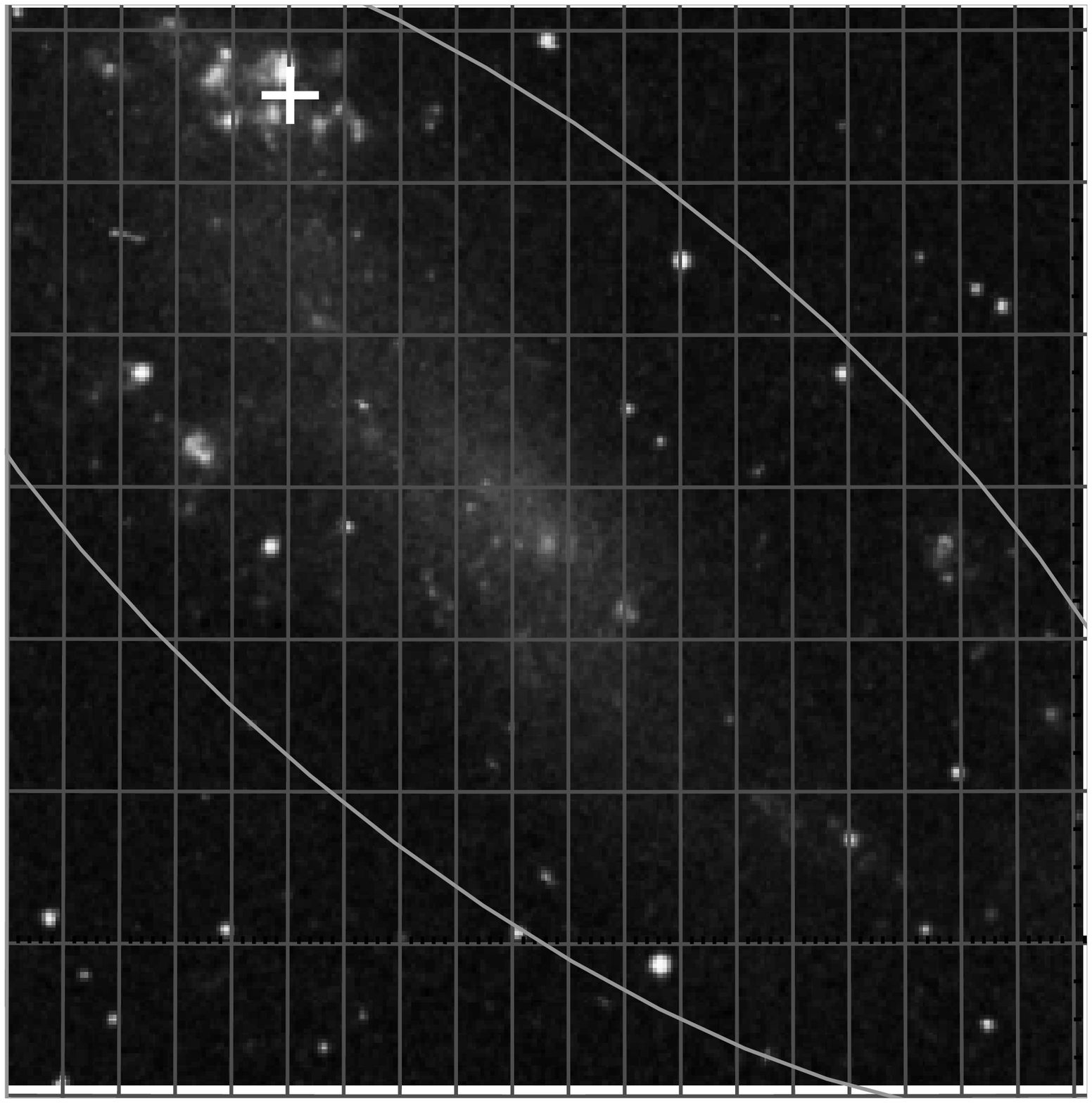}}
\subfigure[\hbox{NGC 1569 \hspace{46pt} Im}]{\includegraphics[scale=0.22]{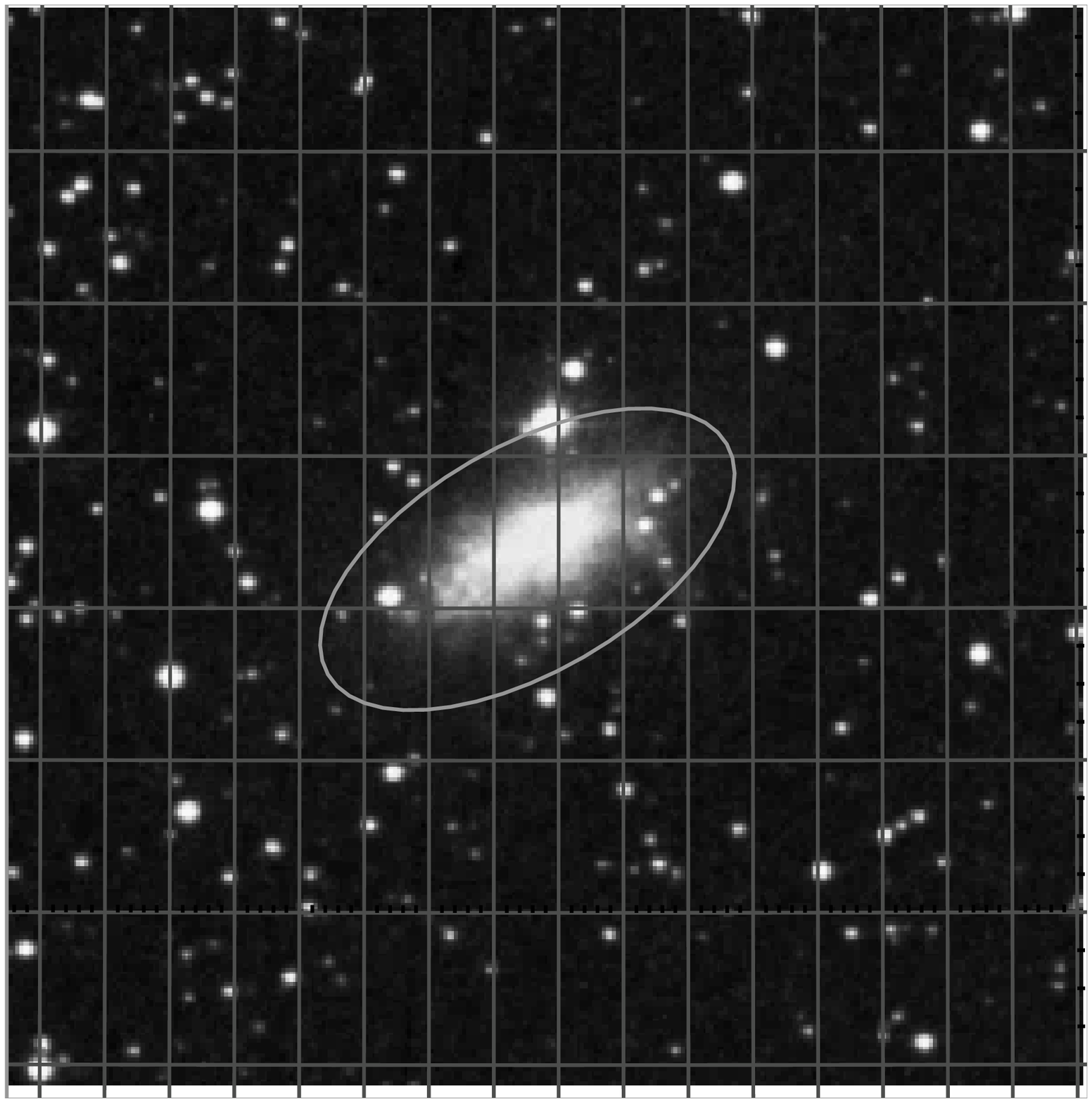}}
\subfigure[\hbox{IZW 18 \hspace{56pt}} ]{\includegraphics[scale=0.22]{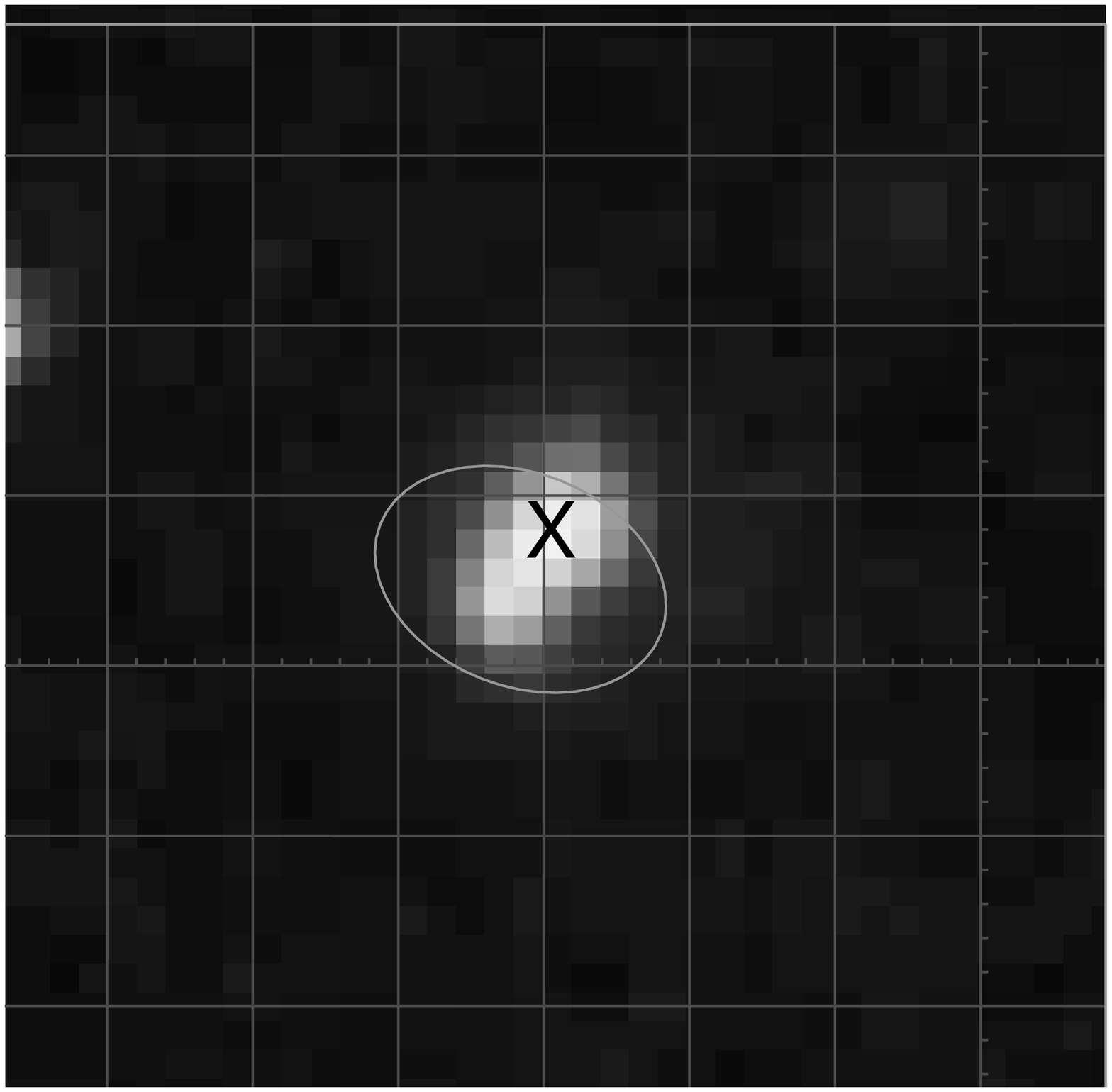}}
\caption{Digital Sky Survey images of each of the galaxies in the sample
with X-ray source detections within the $B$-band 25th magnitude
isophotes superimposed. In all cases, the grid
unit is 1.2$\arcmin$ square, except for IZW\th 18 which has a
grid unit of 0.24\arcmin$\,$ in Right Ascension and 0.18\arcmin$\,$
in Declination, and the image has been magnified by a factor of 8 for
clarity. The ULX (with $L_{\rm x}>L_{10}$) are
indicated with an ``X''; other sources ($L_{\rm x}>L_{5}$)
exceeding the Eddington limit for a neutron star are shown by a
``+''.  Elliptical galaxies are shown at the top, then spiral
galaxies, and finally the irregular galaxy. Except for
NGC\th 253 and IC\th 2574, the detector chips enclosed all of the
region within the isophote; in the latter galaxy the chip
boundaries are shown. In IC\th 5332 and NGC\th 1569, no
detections were made.}
\end{figure*}
\tabcolsep 2.4 mm
\begin{table*}
\begin{center}                           
\caption{The numbers of bright sources detected (see text); $N_{10}$ are
the numbers of ULX. The number of sources with luminosities
greater than ${\rm 2\times 10^{38}}$ erg s$^{-1}$ ($N_2$), ${\rm
5\times 10^{38}}$ erg s$^{-1}$ ($N_5$), ${\rm 10^{39}}$ erg
s$^{-1}$ ($N_{10}$) and ${\rm 10^{40}}$ erg s$^{-1}$ ($N_{100}$)
are shown separately.  Errors take into account both flux and
distance uncertainties. The expected number of background sources
(N$^{bkg}$) for each lower limit luminosity is also shown,
computed at the best distance.
\label{}}
\begin{tabular}{lrrrrrrrrr}
\hline\noalign{\smallskip}
Galaxy & Dist (Mpc)  &  $N_2$ &$N_2^{\rm bkg}$ &
$N_5$  & $N_5^{\rm bkg}$ & $N_{10}$ &$N_{10}^{\rm bkg}$  &
 $N_{100}$ & $N_{100}^{\rm bkg}$\\
\hline\noalign{\smallskip}
NGC\th 4636 &$14.7\pm 0.9$ &$20\pm 5$ &$3.7$ &$3\pm 0.3$ &$1.5$ &$<0.2$ &$0.74$ & 0 & 0.074 \\
NGC\th 1132 &$69\pm 5$ &$2$ &$9.9$ &$2$ &$4.0$ &$2$ &$2.0$ & 1  & 0.2 \\
NGC\th 4697 &$11.7\pm 0.8$ &$14\pm 2$ &$2.8$ &$4\pm 1.2$ &$1.1$ &$<0.06$ &$0.57$ & 0 & 0.06 \\
NGC\th 1399 &$20\pm 1$ &$74\pm 5$ &$11.$ &$26\pm 3$ &$4.5$ &$9\pm 2.5$ &$2.3$ & 0 &0.23  \\
NGC\th 1291 &$12\pm 3$ &$21\pm 10$ &$7.3$ &$8\pm 4.6$ &$3.0$ &$3\pm 1.8$ &$1.5$ & 1 & 0.15 \\
NGC\th 2681 &$17\pm 3$ &$9\pm 1.2$ &$2.2$ &$5\pm 1.5$ &$0.87$ &$3\pm 0.9$ &$0.43$ &0& 0.04 \\
NGC\th 253 &$3.1\pm 0.7$ &$10\pm 3$ &$1.1$ &$7\pm 2.2$ &$0.45$ &$3\pm 2.0$ &$0.22$&0& 0.22 \\
NGC\th 3184 &$7.2\pm 1.7$ &$3\pm 2.8$ &$1.6$ &$2\pm 1.1$ &$0.65$ &$<0.5$ &$0.33$&0& 0.03 \\
NGC\th 4631 &$3.5\pm 0.3$ &$2\pm 0.4$ &$0.31$ &$2\pm 0.2$ &$0.12$ &$1\pm 0.3$ &$0.062$ &0& ${6\times10^{-3}}$ \\
IC\th 5332 &$4.0\pm 1.0$ &$<0.5$ &$0.47$ &$<0.2$ &$0.19$ &$<0.06$ &$0.094$ &0& ${9\times 10^{-3}}$ \\
IC\th 2574 &$3.6\pm 0.3$ &$2\pm 0.5$ &$0.50$ &$1\pm 0.2$ &$0.20$ &$<0.06$ &$0.11$ &0& 0.011 \\
NGC\th 1569 &$1.7\pm 0.2$ &$<0.1$ &$9.7\times 10^{-3}$ &$0$ &$4.2\times 10^{-3}$ &$0$ &$2.3\times 10^{-3}$ &0& $2\times 10^{-4}$ \\
IZW\th 18 &$13\pm 2$ &$1\pm 0.01$ &$6.0\times 10^{-3}$ &$1\pm 0.06$ &$2.4\times 10^{-3}$ &$1\pm 0.3$ &$1.2\times 10^{-3}$&0& $10^{-4}$ \\
\hline\noalign{\smallskip}
Total & \ldots & $158 \pm 13$ & 40.9 & $61\pm 6$ & 16.6 & $22\pm 4$ & 8.40 & 2 & 1.0 \\
\hline\noalign{\smallskip}
\end{tabular}
\end{center}
\end{table*}
\hskip - 5mm
marked ``-'' which would fall below $L_{10}$ at their lower
limit.  The lower panel shows sources exceeding $L_5$ which would
join the ULX detections if given their upper error limit
luminosity (marked ``+'').  Table 5 similarly lists the extended
sources detected.

Next, we estimate the completeness of our source samples, for
each galaxy, i.e. the probability that a source of given luminosity
will be detected by the detection algorithm. This requires firstly
the estimated number of counts of the source in the image, this requiring
a flux to counts conversion factor. We chose a conservatively low factor,
i.e. an estimate erring on the low side of the counts obtained, 
by adopting a simple absorbed power law model having a 
photon index fixed at 1.0, i.e. a smaller value than typical of spectral
fitting results.
The column density was fixed at the average
value measured in the sources detected.  Averaging the threshold
of the ``correlation parameter'' used by {\sc wavdetect} to
identify sources in the vicinity of each of our detections and
assuming a Gaussian distribution of the correlation parameter
(Freeman et al. 2002), we can estimate the fraction of the
complete number of sources above each limit that will be detected
by the algorithm.  This is $>$97\% for $L_5$ sources (except in
NGC\th 1132), and $>94$\% for $L_2$ sources (except for NGC\th
1132, NGC\th 1399, NGC\th 1291 and IZW\th 18). In these objects,
the number detected represents a lower limit.

As the Eddington limit for a 1.4 M$_{\sun}$ neutron star is
between $\rm {2-4\times 10^{38}}$ erg s$^{-1}$, depending on
composition, opacity and gravitational redshift (Paczy\'nski
1983), luminosities smaller than $L_5$ do not exceed the
Eddington limit substantially, and may be neutron star
binaries. Only a very small number of Galactic neutron
star binaries ($<$ 1\%) are known to ever exceed the Eddington
limit: the bright LMXB \hbox{GX\th 5-1} has been observed with
$L_{\rm x}\sim 4-5\times 10^{38}$ erg~s$^{-1}$ (Church \& Ba\l
ucinska-Church 2001; Christian \& Swank 1997). In Sco\ X-1, the
total luminosity often exceeds $L_{10}$. In Table 3 we show the
detections of objects brighter than various thresholds. However,
in the rest of the paper we concentrate on the ULX ($L_{\rm x}$
$>$ $L_{10}$), and do not consider at all objects fainter than
$L_5$ thus excluding most neutron star binaries.  The number of
ULX detected is 22, and these exceed the Eddington limit for a 10
M$_{\sun}$ black hole. There are 39 other sources in the
luminosity range $5-10\times 10^{38}$ erg~s$^{-1}$ which may be
expected to consist mostly of Black Hole Binaries (BHB).

\tabcolsep 3mm
\begin{table*}
\begin{center}
\caption{Upper table: the ULX detections (including sources in
which the errors would allow the luminosity to fall just below
$L_{10}$, marked ``-''); lower table: sources with luminosity
close to $L_{10}$ in which the luminosity errors would allow to
join the ULX category (+ cases).  The {\it Chandra} name includes
the right ascension and declination of each source (epoch 2000); the total
count, count rate and offset from the centre of the galaxy in
arcseconds are shown.}
\begin{minipage}{150mm}
\begin{tabular}{llrrrrr}
\hline\noalign{\smallskip}
Source & name & count &
rate & $L(0.3-7.0)$ & ${\rm \Delta R}$ &$\pm$\\
&&&10$^{-3}$ c s$^{-1}$& $10^{38}$ erg s$^{-1}$& \arcsec\\
\hline\noalign{\smallskip}
\multicolumn{7}{l}{ULX}\\
\hline\noalign{\smallskip}
\multicolumn{7}{l}{NGC\th 1132}\\
PSX-1 & CXOU J025251.4-011631  &$21$ & $1.6$ &$120^{+70}_{-50}$  &$4$  & \\ 
PSX-2 & CXOU J025249.4-011620  &$10$ & $0.76$ &$48^{+37}_{-17}$  &$36$  &  \\
\multicolumn{7}{l}{NGC\th 1399}\\
PSX-1 &CXOU J033832.6-352705   &$700$ & 13.0 &$41.\pm 4.$  &$54.$  &  \\
PSX-2  &CXOU J033831.8-352604  &$570$ & 10.0  &$22.\pm 2.$  &$68.$  &  \\
PSX-3 & CXOU J033820.1-352446  &$410$ & $7.3$  &$15.\pm 2.$  &$190$  &  \\
PSX-4  &  CXOU J033827.6-352648&$520$ & $9.3$  &$14.\pm 2.$  &$22$  &  \\
PSX-5  &CXOU J033821.9-352928  &$190$ & $3.4$ &$13.\pm 2.$  &$180$  &  \\
PSX-6 & CXOU J033829.7-352504  &$230$ & $4.1$  &$13.^{+4.}_{-3.}$  &$110$  &  \\
PSX-7  &  CXOU J033825.2-352522&$340$ & $6.1$  &$11.\pm 1.$  &$110$  &  \\
PSX-8  & CXOU J033832.3-352710 &$120$ & $2.1$ &$11.^{+5.}_{-4.}$  &$51$  & -- \\
PSX-9  & CXOU J033833.1-352731 &$230$ & $4.1$ &$11.\pm 2.$  &$70$  &  -- \\
\multicolumn{7}{l}{NGC\th 1291}\\
PSX-1  & CXOU J031718.6-410629&$390$  & 17.0 &$130^{+140}_{-70}$  &$4$  &   \\ 
PSX-2  &CXOU J031702.5-410714 &$360$ & 16.0  &$18.\pm 2.$  &$230$  &  \\ 
PSX-3  &CXOU J031713.8-411035 &$310$  & 13.0 &$17.\pm 3.$  &$250$  &  \\ 
\multicolumn{7}{l}{NGC\th 2681}\\
PSX-1  &  CXOU J085335.7+511917 &$210$ & $2.7$ &$44.^{+9.}_{-8.}$  &$46$  &  \\ 
PSX-2 & CXOU J085324.4+511819  &$45$ &  $0.6$ &$18.^{+8.}_{-6.}$  &$130$  &  \\ 
PSX-3  & CXOU J085333.7+511930 &$380$ & $4.9$ &$12.\pm 1.$  &$37$  &  \\ 
\multicolumn{7}{l}{NGC\th 253}\\
PSX-1  & CXOU J004734.0-251637 &$650$ & 46.0 &$12.\pm 1.$  &$44$  &  \\ 
PSX-2 & CXOU J004733.0-251749  &$1100$  & 79.0 &$12.1\pm 0.8$  &$31$  &  \\ 
PSX-3  & CXOU J004733.4-251722 &$810$ & 58.0 &$11.^{+5.}_{-2.}$  &$6$  & --  \\
\multicolumn{7}{l}{NGC\th 4631}\\
PSX-1   & CXOU J124211.1+323236&$1100$ & 19.0 &$13.4\pm 0.9$  &$52$  &  \\ 
\multicolumn{7}{l}{IZW\th 18}\\
PSX-1 &   CXOU J093402.0+551428&$390$ & 13.0 &$16.\pm 2.$  &$5$  &  \\ 
\hline\noalign{\smallskip}
\multicolumn{7}{l}{Possible ULX detections}\\
\hline\noalign{\smallskip}
\multicolumn{7}{l}{NGC\th 4636}\\
PSX-1  & CXOU J124249.1+024046 &$140$ & $2.7$  &$7.6^{+7.3}_{-5.5}$  &$33$  & + \\ 
PSX-3  & CXOU J124255.8+024035 &$69$ & $1.3$ &$6.7^{+4.1}_{-2.6}$  &$99$  & + \\
\multicolumn{7}{l}{NGC\th 1399}\\
PSX-10 &  CXOU J033828.6-352724&$58$ & $1.0$ &$9.6^{+29.}_{-6.8}$  &$27$  & +  \\
PSX-11  & CXOU J033826.5-352732 &$150$ & $2.7$ &$9.5\pm 2.1$  &$50.$  &  + \\
PSX-12  & CXOU J033832.8-352658 &$140$ & $2.5$ &$9.2\pm 2.9$  &$56.$  &  + \\
PSX-13  &  CXOU J033827.9-352747&$78$ & $1.4$ &$9.0^{+8.9}_{-4.4}$  &$51$  & +  \\
PSX-14  &CXOU J033832.3-352702 &$98$ & $1.8$ &$8.8^{+5.8}_{-3.5}$  &$50$  & + \\
PSX-15 & CXOU J033815.4-352628 &$210$ & $3.7$ &$8.5\pm 1.9$  &$200$  &  + \\
PSX-17  & CXOU J033833.8-352658 &$71$ & $1.3$ &$7.3^{+11.}_{-4.1}$  &$71$  & +  \\
PSX-18  & CXOU J033827.8-352750  &$110$ & $2.0$ &$7.3^{+16.}_{-2.4}$  &$55$  &  +\\
PSX-22 &  CXOU J033830.4-352430 &$90$& $1.6$  &$6.4^{+9.2}_{-3.7}$  &$150$  & +  \\
\multicolumn{7}{l}{NGC\th 1291}\\
PSX-4  & CXOU J031712.1-410438&$27$  & $1.2$  &$7.8\pm 5.3$  &$140$  & +  \\ 
\multicolumn{7}{l}{NGC\th 2681}\\
PSX-4  & CXOU J085336.4+511727 &$59$ & $0.8$ &$8.2\pm 2.9$  &$99$  & + \\ 
\hline\noalign{\smallskip}
\end{tabular}
\end{minipage}
\end{center}
\end{table*}

\tabcolsep 3mm
\begin{table*}                  
\addtocounter{table}{-1}
\begin{center}
\begin{minipage}{150 mm}
\caption{ contd. Other super-Eddington detections ($L_{\rm x}$ $>$ $L_5$).
In none of these sources do the luminosity errors allow the source to join the
ULX category}
\label{}
\begin{tabular}{llrrrrr}
\hline\noalign{\smallskip}
Source & name & count  & rate &
$L(0.3-7.0)$ & ${\rm \Delta R}$\\
&&&10$^{-3}$ c s$^{-1}$& $10^{38}$ erg s$^{-1}$& \arcsec\\
\hline\noalign{\smallskip}
\multicolumn{7}{l}{Other super-Eddington detections}\\
\hline\noalign{\smallskip}
\multicolumn{7}{l}{NGC\th 4636}\\
PSX-2  & CXOU J124251.4+023835 &$220$ & $4.2$ &$7.0^{+1.5}_{-2.0}$  &$160$  &  \\
\multicolumn{7}{l}{NGC\th 4697}\\
PSX-1  & CXOU J124846.8-054854 &$350$ & $9.0$ &$7.0\pm 1.2$  &$170$  &  \\
PSX-2  & CXOU J124833.2-054742 &$200$ & $5.1$ &$6.4\pm 1.1$  &$45$  &  \\
PSX-3  &CXOU J124836.7-054732  &$130$ & $3.3$ &$5.3\pm 1.2$  &$32$  &   \\
PSX-4  & CXOU J124839.3-054808 &$190$ & $4.9$ &$5.2\pm 0.8$  &$51$  &   \\
\multicolumn{7}{l}{NGC\th 1399}\\
PSX-16  &CXOU J033831.9-352649 &$170$ & $3.0$ &$7.5\pm 1.1$  &$44.$  &  \\
PSX-19  &  CXOU J033831.3-352411 &$200$ & $3.6$ &$7.3\pm 1.6$  &$170$  &  \\
PSX-20  &  CXOU J033828.9-352602 &$130$ & $2.3$ &$7.1\pm 1.7$  &$56$  &  \\
PSX-21  & CXOU J033836.8-352747 &$200$  & $3.6$&$7.0\pm 1.0$  &$130$  &  \\
PSX-23  &  CXOU J033836.3-352809&$61$ & $1.1$ &$6.4\pm 2.2$  &$130$  &  \\
PSX-24  &CXOU J033827.2-352601  &$110$ & $2.0$ &$5.2\pm 1.4$  &$63$  & \\
PSX-25  & CXOU J033831.6-352600 &$160$ & $2.9$ &$5.1\pm 0.9$  &$70$  &  \\
PSX-26 & CXOU J033816.5-352745 &$60$ & $1.1$ &$5.0^{+3.5}_{-2.3}$  &$190$  &  \\
\multicolumn{7}{l}{NGC\th 1291}\\
PSX-5  &CXOU J031658.7-410406 &$67$ & $2.9$ &$6.4^{+2.4}_{-1.9}$  &$310$  & \\ 
PSX-6  & CXOU J031718.9-410628&$79$ & $3.4$ &$5.4\pm 1.7$  &$20$  & \\ 
PSX-7  & CXOU J031721.7-410653&$83$  & $3.6$ &$5.1\pm 1.4$  &$65$  & \\ 
PSX-8  &CXOU J031719.3-410615 &$86$ & $3.7$ &$5.0\pm 1.4$  &$29$  &  \\ 
\multicolumn{7}{l}{NGC\th 2681}\\
PSX-5  & CXOU J085333.5+511852 &$140$ & $1.8$  &$5.1\pm 1.2$  &$6$  & \\ 
\multicolumn{7}{l}{NGC\th 253}\\
PSX-4 &  CXOU J004717.6-251811 &$770$ & 55.0 &$7.8\pm 0.6$  &$240$  &  \\ 
PSX-5 & CXOU J004722.6-252051  &$880$ & 63.0  &$6.5\pm 0.5$  &$260$  &  \\ 
PSX-6 & CXOU J004732.1-251721  &$140$ & 10.0  &$5.7\pm 1.5$  &$16$  &  \\ 
PSX-7  & CXOU J004735.3-251512 &$490$ & 35.0 &$5.2\pm 0.5$  &$130$  &   \\ 
\multicolumn{7}{l}{NGC\th 3184}\\
PSX-1  & CXOU J101823.0+412742 &$460$ & 11.0 &$7.1\pm 0.8$  &$160$  &  \\ 
PSX-2 & CXOU J101812.0+412421  &$420$  & 10.0 &$5.0\pm 0.6$  &$100$  &  \\ 
\multicolumn{7}{l}{NGC\th 4631}\\
PSX-2  & CXOU J124155.6+323217 &$3200$ & 54.0  &$7.2\pm 0.3$  &$180$  &  \\ 
\multicolumn{7}{l}{IC\th 2574}\\
PSX-1  & CXOU J102843.1+682816 &$440$  &51.0 &$6.6\pm 0.8$  &$350$  &  \\ 
\hline\noalign{\smallskip}
\end{tabular}
\end{minipage}
\end{center}
\end{table*}                                    
\tabcolsep 2.8 mm
\begin{table*}
\begin{center}
\begin{minipage}{170 mm}
\caption{The extended source $L_{\rm 5}$ detections}
\begin{tabular}{lllrrrrrr}
\hline\noalign{\smallskip}
&Galaxy&source & $\alpha$ & $\delta$ & count  & rate &
$L(0.3-7.0)$ & ${\rm \Delta R }$\\
&&& h m s &$\degmark \arcmin\;\; \arcsec$&&10$^{-3}$ c s$^{-1}$& $10^{38}$ erg s$^{-1}$& \arcsec\\
\hline\noalign{\smallskip}
&NGC\th 4636\\
&&ESX-1  &  12 42 50.2 &2 41 15 &$620$  & 12.0  &$22^{+49}_{-13}$  &$5.8$  \\
&&ESX-2  &  12 42 49.8 &2 41 15 &$900$ & 17.0  &$18^{+6}_{-5}$  &$1.7$  \\
&&ESX-3  &  12 42 49.7 &2 41 11 &$1100$ & 21.0  &$16^{+4}_{-2}$  &$6.3$  \\
&&ESX-4  &  12 42 50.2 &2 41 10 &$400$ & $7.7$ &$7.5\pm 2.9$  &$9.4$  \\
&NGC\th 1132\\
&&ESX-1  &  2 52 51.8 &-1 16 28 &$100$ & $7.7$  &$140\pm 30$  &$5.5$  \\
&NGC\th 4697\\
&&ESX-1  &  12 48 35.9 &-5 48 3 &$180$ & $4.6$  &$4.5\pm 0.9$  &$1.0$  \\
&NGC\th 1399\\
&&ESX-1  &  3 38 29.0 &-35 27 1 &$11000$  & 200.0 &$760^{+110}_{-80}$  &$3.4$  \\
&NGC\th 2681\\
&&ESX-1  &  8 53 32.7 &51 18 49 &$900$ & 12.0 &$19\pm 2$  &$6.8$  \\
&NGC\th 253\\
&&ESX-1 &   0 47 33.3 &-25 17 17 &$410$ & 29.0 &$10^{+3}_{-2}$  &$2.4$  \\
&&ESX-2 &   0 47 33.0 &-25 17 20 &$600$  & 43.0 &$10\pm 1$&$2.1$  \\
\hline\noalign{\smallskip}
\end{tabular}
\end{minipage}
\end{center}
\end{table*}


\subsection{Radial intensity profiles}

\begin{figure*}
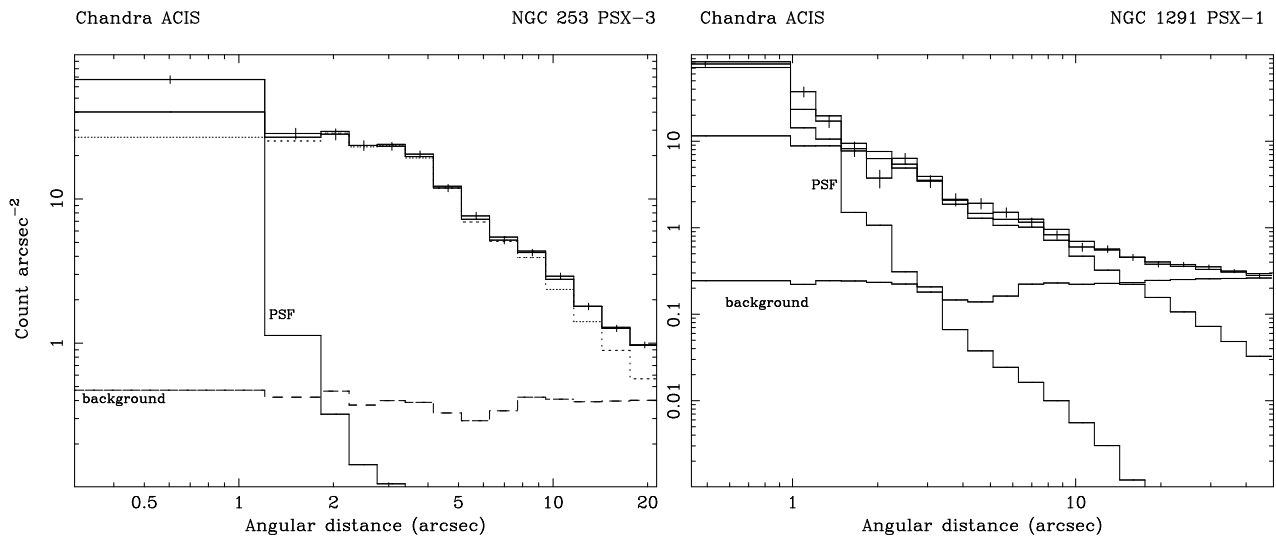

\includegraphics[width=70mm,angle=270]{f2a}             
\includegraphics[width=70mm,angle=270]{f2b}            
\caption{Point sources embedded in extended emission.  Radial
brightness profiles of NGC\th 253~PSX-3 (left panel) and
NGC\th~1291~PSX-1 (right panel), showing the clearly extended
emission. This has been parameterized in both cases using King
models.  The contributions of individual components in the
fitting (background, PSF and King model) are shown.
\label{rpsfplot}}
\end{figure*}

The excellent spatial resolution of {\em Chandra} (Weisskopf et
al. 2002) makes it possible to test for spatial extension of the
ULX.  Accordingly, radial intensity profiles for each detected
bright source were extracted, to search for excesses above the
point-spread function. Since the PSF width depends on the
off-axis angle and the source spectrum, PSFs were generated for
each source based on the mean detected photon energy using the
{\sc ciao} tool {\sc mkpsf}. Source data were extracted in 18
logarithmically-spaced radial bins in the total band 0.3--7.0 keV
and rebinned to ensure a minimum of 20 counts per bin, and
compared with the PSF. This was done using dedicated software in
which it was tested whether a point source model fits the data
well. It was also possible to fit extended source models such as
a simple King model. The background was assumed to be constant
over the radial distance in the radial profiles, and all point
sources detected in the neighbourhood of the source being
modelled were excluded from the background data.  In the majority
of sources, there was no evidence of extended nature.  The
extended sources are not included in Table 4, or in discussion of
ULX, but are shown separately in Table 5.

In the case of the detection NGC\th 253 PSX-3, the radial profile
shown in Fig. 2 (left panel) consists of clear extended emission
many times wider than the PSF, together with excess intensity at
the centre with a width comparable to the PSF, suggesting a
point-source, and so this source in included in Table 4.  In
NGC\th~253 and NGC\th~4636, there are known to be regions of
substantial, diffuse galaxy emission exhibiting complicated
structure (Strickland et al. 2000; Jones et al. 2002), so that it
is difficult to test if sources close to the centres are real or
concentrations of structured emission.  Similarly, where this
extended emission has an erratic radial profile that is not
easily fitted, it is difficult to determine whether sources
superimposed upon it are broader than the PSF or genuine
point-sources.  In such ambiguous cases the objects are included
in Table 5 and assumed to be extended.

In five galaxies, detections were made within a few arcseconds of
the galaxy centres where there may be diffuse emission, so that
the sources may not be point-like. The radial profiles of four of
these sources (NGC\th 1132 ESX-1, NGC\th 1399 ESX-1, NGC\th 2681
ESX-1 and NGC\th 4697 \hbox{ESX-1}) 
could be fitted by King models, and
so the sources are shown in Table 5 of extended sources, also
containing extended sources offset from the centres as revealed
by their radial profiles.  The fitting results allowed the
extended emission to be subtracted in fitting the radial profiles
of other sources in these five galaxies, and when this was done,
no other source within these galaxies proved to be extended.  In
the case of NGC\th 1291, a point-source PSX-1 was detected at the
galaxy centre, superimposed on extended emission (see Fig. 2,
right panel); this is discussed in Sect. 5.3.

\subsection{Hardness ratios}

As a simple method of comparing the sources, hardness ratios were
obtained for all sources in Table 4 based on three energy bands:
a soft band (S): 0.3--1.5 keV, a medium band (M): 1.5--3.0 keV
and a hard band (H): 3.0--5.0 keV.  Hardness ratios were defined
as HR1 = (S-M)/(S+M) and HR2 = (M-H)/(M+H).  The HR1 ratio is
more sensitive to changes in column density, while the HR2 ratio
is more sensitive to the power law index.  With this definition,
a harder source implies a more negative value of HR2.  A
colour-colour diagram (HR2 {\it versus} HR1) is shown for the
detected sources in each galaxy in Fig. 3, plotting each source
as a point with error bars derived from the Poisson errors of the
intensities in each energy band.

\begin{figure*}                 
\includegraphics[scale=0.8]{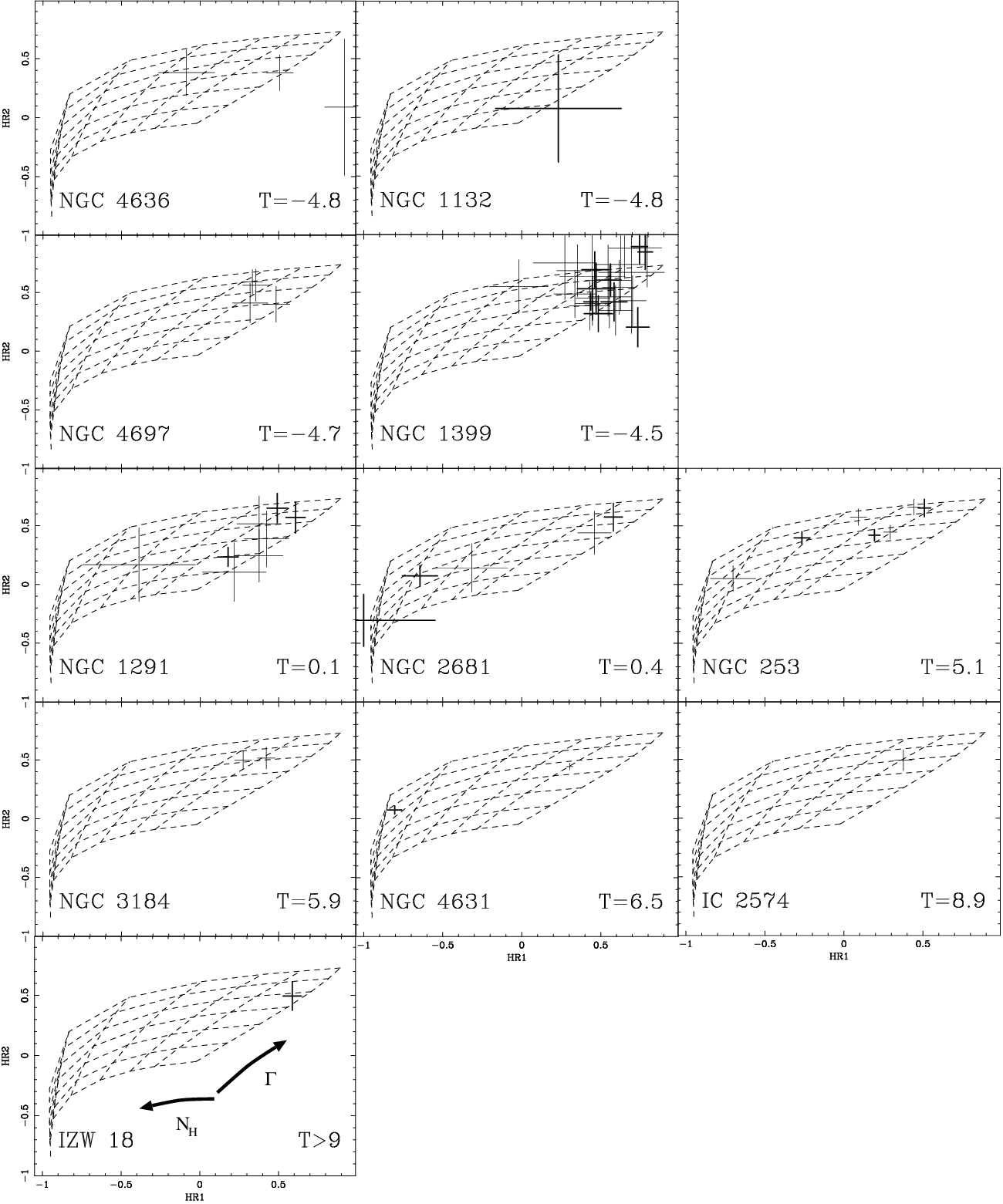} 
\caption{ 
X-ray colour-colour plot for sources with $L_{\rm x} > 5\times
10^{38}$ erg s$^{-1}$ in each galaxy, ordered by morphological
parameter $T$. ULX sources are shown with darker lines; other
super-Eddington sources are shown with fainter lines.
Each panel shows a grid of power law index
$\Gamma$ - $N_{\rm H}$ values obtained by simulating ACIS data
for an absorbed power law spectral model (see text). $\Gamma$
increases vertically with values between 0 -- 3.0 in steps of
0.5, and $N_{\rm H}$ increases to the left with values 0.0, 0.25,
0.5, 1.0, 2.0, 4.0 and 8.0 ${\rm \times 10^{22} cm^{-2}}$.  In
IC\th 5332 and NGC\th 1569 no super-Eddington sources were
detected, so they are omitted.}
\end{figure*}
To investigate quantitatively the sensitivity of the ratios of
these bands to spectral changes, ACIS data were simulated for a
simple absorbed power law spectral model for a range of column
densities $N_{\rm H}$ and power law indices $\Gamma$. Count rates
in each of the energy bands were calculated and points on the
(S/M, M/H) colour-colour diagram found, so giving a
$\Gamma$/$N_{\rm H}$ grid (c.f. Kim et al. 1992). This grid is
superimposed on the colour-colour diagram for the sources found
in each galaxy in Fig. 3. (The $\Gamma$ scale is vertical, the
$N_{\rm H}$ scale horizontal, with increasing values to the top
and left.)

Although the spectra of LMXB, HMXB or BHB cannot generally be
described by this simple one-component model, an absorbed
power law model can be fitted to X-ray binary spectra as a method
for gauging the steepness of the decrease to high energies, even
if the quality of fit may be poor. For example, Cyg\th X-1 (a
BHB) in the High State has a very soft spectrum because of the
strong disc blackbody component (Dotani et al. 1997). Although
the power law index of the Comptonized emission varies from
$\sim$1.7 to 2.0 as the source moves from the Low State to the
High state, fitting the one-component model to High State data
will yield an apparent $\Gamma$ of $\sim$3.5, compared with
$\sim$1.9 in the Low State.  We have carried out simulations
based on previous analysis of Galactic Black Hole binaries and
Low Mass X-ray Binaries, and results are shown in Fig. 4. The
simulation based on the Cyg\th X-1. High State data of Ba\l
uci\'nska-Church et al. (1998) confirms a location on the
colour-colour diagram at a grid value with $\Gamma$ = 3.5, so
that sources located at similar positions in Fig. 3 may be Black
Hole Binaries in the High State. Power law indices between
2.5--3.5 may indicate black hole nature; however given the errors
of points in Fig. 3, it would be unsafe to associate black hole
nature except with values $>$3.0.  There were five such objects,
including two ULX: NGC\th 1399 PSX-1 and PSX-2, plus NGC\th 1399
PSX-15 (luminosity marginally below $\rm {10^{39}}$ erg
s$^{-1}$), and PSX-16 and PSX-26, having $L_{\rm x}$ $>$ $L_5$.
Note that large values of $\Gamma$ may also result from soft
thermal spectra, for example those of SNR or supersoft sources.

Colour-colour diagrams however cannot distinguish a BHB in the
Low State from a LMXB, although most LMXB should have been
excluded from the sample defined with \hbox{$L_{\rm x} > 5\times
10^{38}$ erg s$^{-1}$.}  Moreover, it would not be possible to
distinguish different classes of LMXB; for example, the flaring
branch in Z-track sources is distinguished from the normal branch
by flaring taking place above 7 keV (Ba\l uci\'nska-Church et
al. 2001), but tests show that this behaviour would not cause
any movement in the present colour-colour diagram. However, the
diagram does allow comparisons between galaxies, and with galaxy
morphology. It should be noted that fitting the absorbed power law
model to ACIS data was adequate in only half of the sources shown
in Table 6 for which spectral fitting was carried out.

Differences in hardness ratio between galaxies can be seen.  In
the early-type galaxies, the column densities are low:
$\approxlt$ $5\times 10^{21}$ cm$^{-2}$, as expected since early
type galaxies generally have a hot ISM.  Higher column-densities
are seen however, in NGC\th~4636 PSX-3 and NGC\th~2681 PSX-1,
PSX-2 and PSX-4, sources that are not near the nucleus.
Examination of the spectra of these sources reveals evidence of
heavy absorption in each case. This may be intrinsic, for example
if the sources are strongly-absorbed background AGN. We note that
the expected numbers of background objects (Sect. 3) is
consistent with the numbers of heavily-absorbed sources we detect
in elliptical galaxies.  Alternatively, they may be
intrinsically-absorbed objects within the host galaxy. In the
nearly edge-on spiral galaxy NGC\th~253 we see evidence of a
spread of column-densities consistent with the presence of cold
ISM in spirals, reaching a few $\times 10^{22}$ cm$^{-2}$.
Spectral fitting results (Table 6) confirm high values of column
density in the sources in Fig. 3 with large $N_{\rm H}$.

\subsection{Spectral analysis}

We restrict spectral fitting to sources with $L_{\rm x} > 5\times
10^{38}$ erg s$^{-1}$ and for a total number of counts in the
spectrum of $>$ 200, which is a minimum for useful spectral
fitting. This left a sample of 22 sources consisting of 14 ULX
plus 8 other super-Eddington sources ($L_{\rm
x}>L_{5}$). Spectral fitting results are shown in Table 6 for the
ULX (upper panel), and other sources (lower panel).

Spectra were extracted from the elliptical regions provided by
the detection algorithm and background data extracted from annuli
centred on each source from events files from which all point
sources had been excluded. Data below 0.3 keV, where the
calibration becomes uncertain, and above 7.0 keV, where the
particle background exceeds the source count rate, were
ignored. Spectra were grouped to a minimum of 20 counts per bin
to allow use of the ${\chi^{2}}$ statistic. However, the full
energy range could not be used in many cases because of
insufficient counts. Although dependent on the slope of the
spectrum, to have usable data above 5 keV, more than 500 counts
are needed.  Non-standard spectral models were written for use
within {\sc xspec} in which the intensity or normalization of the
model equals the unabsorbed energy flux in the 0.3--7.0 keV band
thus allowing flux errors to be obtained directly as the error of
a single parameter. Simple absorbed power law (P), multi-colour
disc blackbody (D) and Mekal (M) models were tested against the
data, as were two-component combinations of these.  The Mekal
model would be appropriate to optically thin emission with lines
from hot plasma, appropriate for some SNR.  In no cases was a
cut-off power law representation of Comptonization fitted given
that the spectra did not extend beyond 7 keV and use of this
model is likely to result in curvature at $\sim$5 keV actually
due to thermal emission being incorrectly interpreted as a high
energy cut-off. In Table 5, results shown in boldface indicate
that the model highlighted produced an appreciable decrease in
$\chi^2$ compared with other models.

\begin{figure}
\centering
\includegraphics[width=62mm,angle=270]{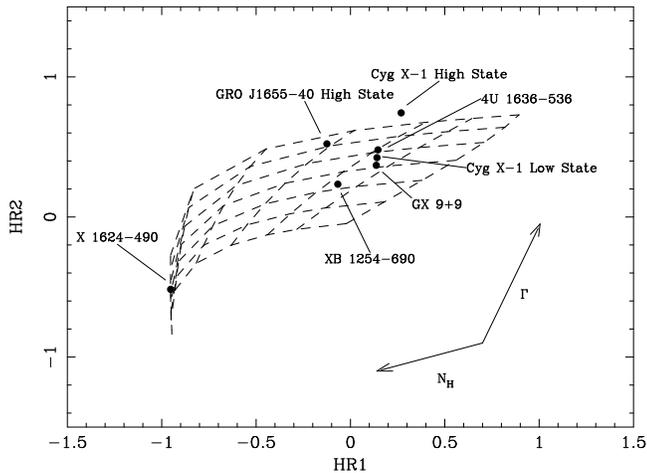} 
\caption{
Calibration of grid of X-ray colour-colour diagram using Galactic
Black Hole Binaries and Low Mass X-ray Binaries (see text).
Cyg\th X-1 and GRO\th J1655-40 are BHB, the other sources are
LMXB.  Simulations were based on the work of Dotani et
al. (1997), Ba\l uci\'nska-Church \& Church (2000), Ba\l
uci\'naka-Church et al. (2001), Church \& Ba\l uci\'nska-Church
(2001), and Smale et al. (2002). Column density and power law
index contours correspond to the same values of these parameters
as in Fig. 3}
\end{figure}

Of the 14 ULX in Table 6, 5 could be fitted by a simple absorbed
power law, although in none of these cases was the power law
photon index large enough ($\Gamma$ $>$ 3.0) to be taken as
evidence for black hole nature (Sect. 3.2). In one source (NGC\th
1399 source PSX-3) a Mekal model was preferred marginally,
offering some evidence for SNR nature. In two sources (NGC\th 253
PSX-1 and NGC\th 4631 PSX-1) an absorbed blackbody model was
preferred although it was not possible to discriminate between
disc blackbody and simple blackbody. These may indicate a black
hole in a high state.

The other 7 sources required two-component models, 5 being fitted
by a blackbody plus power law which would suggest neutron star
binary or black hole binary nature. However, it was not possible
to discriminate between the simple blackbody and disc blackbody
spectral forms and so discriminate between a neutron star binary
and a black hole binary.  IZW\th 18 PSX-1 was best fitted by a
Mekal plus power law, suggesting possible SNR nature, as was
NGC\th 1291 PSX-1, although in this case it was necessary for the
power law to have extra absorption.

Comparison was next made of spectral fitting and hardness ratio
results for ULX. Two ULX were identified from the harness ratio
diagrams with a spectral shape which when fitted by an absorbed
power law model, would give $\Gamma $ $>$ 3.0. These were NGC\th
1399 PSX-1 and PSX-2. Spectral fitting results 
for the second of
these had an actual power law photon index of 2.5-3.5 as part of
a two-component model, adding some support to the idea that it is
a High State BHB.

Of the 8 other (non-ULX) sources shown in the lower panel of
Table 6, exceeding the Eddington limit for a neutron star, 4 can
be fitted by an absorbed power law model, although none of these
had a photon index $>$ 3.0 suggesting black hole nature. The
other 4 were fitted by two-component blackbody plus power law
models consistent with neutron star binary or black hole binary
nature.  In one of these, NGC\th 253 PSX-7, there was evidence
for a steep power law index ($>$ 3.0) strongly suggestive of
black hole binary nature.  The luminosity of this source is $\rm
{5.2\times 10^{38}}$ erg s$^{-1}$.  Very few Galactic neutron
star binaries have been observed with such luminosities, and as
the luminosity is substantially less than the Eddington limit of
a 10 M$_{\sun}$ black hole, the object is likely to be a black
hole binary similar in the high state.
Fig. 5 illustrates higher quality data, showing the best fits to
two sources. NGC\th 1399 PSX-1 (left panel) is fitted by an
absorbed power law model, whereas the second source,

\tabcolsep 0.7mm
\begin{table*}
\begin{minipage}{170mm}
\caption{Spectral fitting results: upper panel: ULX; lower panel:
other super-Eddington sources. Only spectra with $>$ 200 counts
were fitted (see text).  The spectral models used are: P = power
law, B = blackbody, D = disc blackbody, M = Mekal, in all cases
with cool absorber. The M+AP model has additional absorption for
the power law of ${\rm 2.2\pm 1.0\cdot 10^{22}}$ atom cm$^{-2}$.
Fitting results in boldface identify the models
which were significantly preferred.  Although the source NGC\th
1399 PSX-9 had more than 200 counts, fitting had less than 8 dof
and for this reason, sensible fitting was not possible.} 
\label{fitresults}
\begin{tabular}{lllrrlllrrrr}
\hline\noalign{\smallskip}
Obj. & model & $\chi^2$/dof   & $N_{\rm H}^{22}$ &
$kT$ (keV) & $\Gamma$ & Obj. & model & $\chi^2$/dof  & $N_{\rm H}^{22}$ &
$kT$ (keV) & $\Gamma$\\
\hline\noalign{\smallskip}
\multicolumn{12}{l}{ULX ($L_{\rm x}$ $>$ $L_{10}$)} \\
\hline\noalign{\smallskip}
NGC\th 1399 &\ldots &\ldots &\ldots &\ldots &\ldots&\ldots &B+P &$14$/8 &$<$$0.29$ &$0.41^{+2.1}_{-0.38}$ &$<$$3.5$\\
PSX-1 &{\bf P }&${\bf 23/27}$ &${\bf 0.048^{+0.032}_{-0.028}}$ &\ldots & ${\bf 1.5\pm 0.2}$&\ldots &D+P &$13$/8 &$<$$0.71$ &$0.068^{+2.4}_{+10.0}$ 
&$<$$3.5$\\
\ldots &D &$31$/27 &$<$$9.7\times 10^{-3}$ &$1.3\pm 0.2$ &\ldots&\ldots &M+P &$13$/8 &$<$$0.17$ &$1.1^{+19.}_{-1.1}$ &$1.5^{+2.0}_{-0.7}$\\
\ldots &M &$24$/27 &$0.026^{+0.020}_{-0.018}$ &$>$$7.8$ &\ldots&NGC 2681 &\ldots &\ldots &\ldots &\ldots &\ldots\\
PSX-2 &P &$23$/19 &$0.072\pm 0.049$ &\ldots &$2.5\pm 0.4$&PSX-3 &{\bf P }&${\bf 5.4/15}$ &${\bf 0.13^{+0.06}_{-0.05}}$ &\ldots &${\bf 
2.2^{+0.4}_{-0.3}}$\\
\ldots &{B+P}&${19/17}$ &${ <0.11}$ &${ 0.26\pm 0.09}$ &${ 2.5\pm 1.0}$&\ldots &D &$14$/15 &$<$$0.040$ &$0.82\pm 0.18$ &\ldots\\
\ldots &{D+P}&${19/17}$ &${ <0.10}$ &${ 0.41\pm 0.15}$ &${ <3.5}$&\ldots &M &$15$/15 &$0.029^{+0.022}_{-0.019}$ &$4.7^{+2.3}_{-1.1}$ &\ldots\\
\ldots &{\bf M+P}&${\bf 16/17}$ &${\bf 0.058\pm 0.051}$ &${\bf 1.3^{+0.8}_{-0.4}}$ &${\bf 2.6^{+0.6}_{-0.5}}$&NGC 253 &\ldots &\ldots &\ldots &\ldots 
&\ldots\\
PSX-3 &P &$15$/12 &$<$$0.092$ &\ldots &$1.8^{+0.4}_{-0.3}$&PSX-1 &P &$15$/26 &$1.0\pm 0.2$ &\ldots &$2.1\pm 0.3$\\
\ldots &D &$16$/12 &$<$$0.016$ &$0.73\pm 0.15$ &\ldots&\ldots &{\bf D}&${\bf 10/26 }$&${\bf 0.70\pm 0.14}$ &${\bf 1.3\pm 0.2}$  &\ldots\\
\ldots &{\bf M}&${\bf 13/12}$ &${\bf <0.018}$ &${\bf 4.1^{+2.5}_{-1.4}}$ &\ldots&\ldots &M &$14$/26 &$0.87\pm 0.19$ &$5.2^{+2.5}_{-1.4}$ &\ldots\\
PSX-4 &P &$32$/20 &$0.14^{+0.07}_{-0.05}$ &\ldots &$2.8^{+0.5}_{-0.4}$&PSX-2 &P &$51$/43 &$0.36^{+0.06}_{-0.05}$ &\ldots &$1.8\pm 0.2$\\
\ldots &B+P &$20$/18 &$<$$0.079$ &$0.23^{+0.02}_{-0.05}$ &$<$$2.2$&\ldots &B+P &$47$/41 &$0.47\pm 0.11$ &$>$$1.3$ &$2.4\pm 0.6$\\
\ldots &D+P &$23$/18 &$0.046^{+0.063}_{-0.040}$ &$0.34\pm 0.11$ &$<$$2.4$&\ldots &D+P &$48$/41 &$0.31^{+0.39}_{-0.07}$ &$0.56^{+0.28}_{-0.53}$ 
&$<$$3.5$\\
\ldots &M+P &$20$/18 &$0.074^{+0.064}_{-0.049}$ &$1.0\pm 0.3$ &$2.5^{+0.5}_{-0.4}$&\ldots &M+P &$46$/41 &$0.40\pm 0.11$ &$0.61^{+1.4}_{-0.20}$ 
&$1.7\pm 0.2$\\
PSX-7 &{\bf P} &${\bf 13/12}$ &${\bf <0.026}$ &\ldots &${\bf 1.8^{+0.3}_{-0.2}}$&PSX-3 &P &15/20 &$0.40^{+0.40}_{-0.25}$ &\ldots &$2.1\pm 1.6$\\
\ldots &D &$39$/12 &$<$$9.4\times 10^{-3}$ &$0.51\pm 0.16$ &\ldots&\ldots &D &16/20 &$0.22^{+0.30}_{-0.15}$ &$1.2^{+6.3}_{-0.7}$ &\ldots\\
\ldots &M &$31$/12 &$<$$5.9\times 10^{-3}$ &$7.5\pm 6.0$ &\ldots&\ldots &M &15/20 &$0.32^{+0.34}_{-0.16}$ &$>$$1.7$ &\ldots\\
NGC 1291 &\ldots &\ldots &\ldots &\ldots &\ldots&NGC 4631 &\ldots &\ldots &\ldots &\ldots &\ldots\\
PSX-1 &P &$42$/15 &$<$$0.068$ &\ldots &$0.77^{+0.21}_{-0.18}$&PSX-1 &P &$47$/49 &$4.1\pm 0.6$ &\ldots &$2.7\pm 0.3$\\
\ldots &B+P &$32$/13 &$0.25\pm 0.06$ &$1.2^{+0.3}_{-0.2}$ &$>$$2.6$&\ldots &{\bf D}&${\bf 43/49}$ &${\bf 2.9^{+0.4}_{-0.3}}$ &${\bf 1.4\pm 0.2}$ 
&\ldots\\
\ldots &M+P &$28$/13 &$<$$0.20$ &$0.42\pm 0.24$ &$0.59\pm 0.25$&\ldots &M &$61$/49 &$3.3\pm 0.4$ &$3.8^{+1.4}_{-0.8}$ &\ldots\\
\ldots &{\bf M+AP} &{\bf 14/12}&${\bf <0.048}$ &${\bf 0.51\pm 0.15}$ &${\bf 2.3^{+0.6}_{-1.2}}$&IZW 18 &\ldots &\ldots &\ldots &\ldots &\ldots\\
PSX-2 &{\bf P }&${\bf 6.8/12}$ &${\bf 0.13^{+0.08}_{-0.06}}$ &\ldots &${\bf 2.2^{+0.5}_{-0.4}}$&PSX-1 &P &$23$/15 &$0.18^{+0.06}_{-0.05}$ &\ldots 
&$2.3^{+0.4}_{-0.3}$\\
\ldots &D &$11$/12 &$<$$0.066$ &$0.68^{+0.21}_{-0.15}$ &\ldots&\ldots &B+P &$15$/13 &$0.30^{+0.31}_{-0.20}$ &$0.11^{+0.10}_{-0.04}$ 
&$2.0^{+0.7}_{-0.8}$\\
\ldots &M &$10$/12 &$0.031\pm 0.025$ &$4.3^{+2.6}_{-1.2}$ &\ldots&\ldots &D+P &$15$/13 &$0.33^{+0.32}_{-0.20}$ &$0.14^{+0.19}_{-0.05}$ 
&$2.0^{+0.7}_{-1.1}$\\
PSX-3 &P &$15$/10 &$0.11^{+0.10}_{-0.07}$ &\ldots &$1.8^{+0.6}_{-0.5}$&\ldots &{\bf M+P}&${\bf 11/13}$ &${\bf 0.21^{+0.11}_{-0.08}}$ &${\bf 
0.21^{+0.07}_{-0.04}}$ &${\bf 2.1\pm 0.4}$\\
\hline\noalign{\smallskip}
\multicolumn{12}{l}{Super-Eddington sources ($L_{\rm x}$ $>$ $L_5$)} \\
\hline\noalign{\smallskip}
NGC\th 4697 &\ldots &\ldots &\ldots &\ldots &\ldots&\ldots &B+P &$25$/17 &$<$$0.97$ &$0.47^{+0.27}_{-0.13}$ &$<$$2.0$\\
PSX-1 &P &$22$/11 &$<$$0.036$ &\ldots &$1.0\pm 0.2$&\ldots &D+P &$26$/17 &$0.59^{+0.43}_{-0.52}$ &$0.089^{+2.4}_{-0.059}$ &$<$$2.9$\\
\ldots &{\bf B+P} &${\bf 12/9}$ &${\bf <0.039}$ &${\bf 0.68^{+0.10}_{-0.09}}$ &${\bf >2.3}$&\ldots &M+P &$28$/17 &$0.18^{+0.18}_{-0.07}$ 
&$>$$1.0\times 10^{-3}$ &$<$$3.5$\\
\ldots &{\bf M+P}&${\bf 12/9}$ &${\bf <0.22}$ &${\bf 0.026^{+0.017}_{-0.025}}$ &${\bf 1.3\pm 0.4}$&PSX-2 &P &$20$/16 &$0.17^{+0.07}_{-0.06}$ &\ldots 
&$1.9^{+0.4}_{-0.3}$\\
NGC 253 &\ldots &\ldots &\ldots &\ldots &\ldots&\ldots &D &$20$/16 &$0.052\pm 0.037$ &$1.0\pm 0.2$ &\ldots\\
PSX-4 &P &$42$/31 &$0.37^{+0.06}_{-0.05}$ &\ldots &$2.0\pm 0.2$&\ldots &M &$20$/16 &$0.091^{+0.032}_{-0.027}$ &$5.3^{+3.3}_{-1.5}$ &\ldots\\
\ldots &B+P &$22$/29 &$0.90^{+0.32}_{-0.26}$ &$0.098^{+0.026}_{-0.018}$ &$2.1\pm 0.3$&\ldots &B+P &$19$/14 &$<$$0.39$ &$0.79^{+1.7}_{-0.76}$ 
&$2.8^{+0.7}_{-2.5}$\\
\ldots &D+P &$21$/29 &$0.96^{+0.33}_{-0.24}$ &$0.11^{+0.03}_{-0.02}$ &$2.2^{+0.4}_{-0.3}$&\ldots &D+P &$19$/14 &$0.20^{+0.20}_{-0.08}$ &$<$$2.5$ 
&$2.0^{+0.4}_{-0.3}$\\
\ldots &M+P &$24$/29 &$0.38^{+0.08}_{-0.07}$ &$0.65^{+0.21}_{-0.18}$ &$1.8\pm 0.2$&NGC 4631 &\ldots &\ldots &\ldots &\ldots &\ldots\\
PSX-5 &{\bf P}&${\bf 32/35}$ &${\bf 0.23^{+0.05}_{-0.04}}$ &\ldots &${\bf 2.3\pm 0.2}$&PSX-2 &{\bf P }&${\bf 147/115}$ &${\bf 0.23\pm 0.02}$ &\ldots 
&${\bf 1.77\pm 0.09}$\\
\ldots &D &$40$/35 &$0.078\pm 0.025$ &$0.84\pm 0.11$ &\ldots&\ldots &D &$172$/115 &$0.10\pm 0.01$ &$1.31\pm 0.09$ &\ldots\\
\ldots &M &$44$/35 &$0.096\pm 0.019$ &$4.5^{+1.0}_{-0.8}$ &\ldots&\ldots &M &$156$/115 &$0.16\pm 0.01$ &$7.6^{+1.6}_{-0.7}$ &\ldots\\
PSX-7 &P &$30$/20 &$0.45^{+0.11}_{-0.09}$ &\ldots &$2.1\pm 0.3$&IC\th 2574 &\ldots &\ldots &\ldots &\ldots &\ldots\\
\ldots &{\bf B+P}&${\bf 24/18}$ &${\bf 0.13^{+0.43}_{-0.06}}$ &${\bf 0.54\pm 0.10}$ &${\bf <3.5}$&PSX-1 &P &$11$/17 &$0.14^{+0.06}_{-0.05}$ &\ldots 
&$1.7\pm 0.3$\\
\ldots &{D+P}&${\bf 26/18}$ &${ 0.28^{+0.75}_{-0.06}}$ &${ 0.91^{+0.45}_{-0.29}}$ &${ <3.5}$&\ldots &D &$13$/17 &$0.042^{+0.035}_{-0.029}$ 
&$1.2^{+0.3}_{-0.2}$ &\ldots\\
NGC 3184 &\ldots &\ldots &\ldots &\ldots &\ldots&\ldots &M &$11$/17 &$0.095^{+0.034}_{-0.029}$ &$7.0^{+7.3}_{-2.2}$ &\ldots\\
PSX-1 &P &$28$/19 &$0.20^{+0.07}_{-0.06}$ &\ldots &$1.7\pm 0.3$&&&&&&\\
\hline\noalign{\smallskip}
\end{tabular}
\end{minipage}
\end{table*}
\noindent
NGC\th 253 PSX-4 is well-fitted by a two-component model 
consisting of either a simple blackbody or a disc blackbody 
plus a power law.

\section{Search for correlation of ULX numbers with stellar mass and star-forming mass}

We have investigated whether correlations exist between the numbers of ULX
($N_{10}$) and the numbers of super-Eddington sources ($N_5$) with the stellar
masses of the galaxies, and with the mass of star-forming regions in the galaxies.
We used for the first test the near-infrared {\it H}-band fluxes accepted as
a good measure of stellar mass, and for the second test, we used the 60 $\mu$m
fluxes. In both cases, data were obtained from the 
{\it NASA Extragalactic Database} (NED) which
also provided angular diameters for the galaxies in the form $\theta_1$ and $\theta_2$
each given in arcminutes. For each galaxy, we used the total {\it H}-band flux $f$ in 
Janskys from the photometric measurements listed in the database. We adopted
the procedure of taking $f/ \theta_1\cdot\theta_2$ as a measure of the
surface brightness of the {\it H}-band luminosity of each galaxy. The number
counts $N_{10}$ and $N_5$ were converted to counts per unit area by dividing
by the projected area of each galaxy $d^2 \theta_1\cdot\theta_2$, where $d$
is the galaxy distance. It is well-known that this procedure avoids possible
biases due to varying sizes of the galaxies. {\it H}-band fluxes were available
from the database for all of the galaxies in the sample, except IZW\th 18.
Similarly, there was no 60 $\mu$m flux for IZW\th 18 and only upper limits for
NGC\th 1399 and NGC\th 1132. These upper limits were not included in Fig. 7
in which we test for correlation of number counts with 60 $\mu$m flux.
We use the $N_{10}$ and $N_5$ number counts listed in Table 3 which include
5 upper limits for $N_{10}$ and 1 upper limit for $N_5$; these upper limits
are plotted in Figs. 6 and 7 but are excluded from the correlation analysis.
In NGC\th 1569, no sources were detected.

In Fig. 6 we show the variation of $N_{10}$ and $N_5$ per unit area with
{\it H}-band surface brightness as an indicator of the stellar mass of each galaxy.
Neither of these suggests a correlation with stellar mass.
The modified Kendal's-$\tau$ statistic demonstrated a 88.1\%
probability of {\it no} correlation between $N_{10}$ and stellar mass, and a 78.8\%
probability of {\it no} correlation between $N_{5}$ and stellar mass.
In Fig. 7 we show the corresponding plots for 60 $\mu$m surface brightness.
The $N_5$ counts reveal a strong correlation with the infrared surface brightness
with a correlation coefficient of 0.71 and probability of 98.7\% of correlation.
In the case of the $N_{10}$ data, there are only 4 points that are not
upper limits, but these also can be seen to be correlated with 60 $\mu$m flux, 
although the formal probability of 82.6\% may be regarded as insufficiently
high to {\it prove} the correlation.

\begin{figure*}
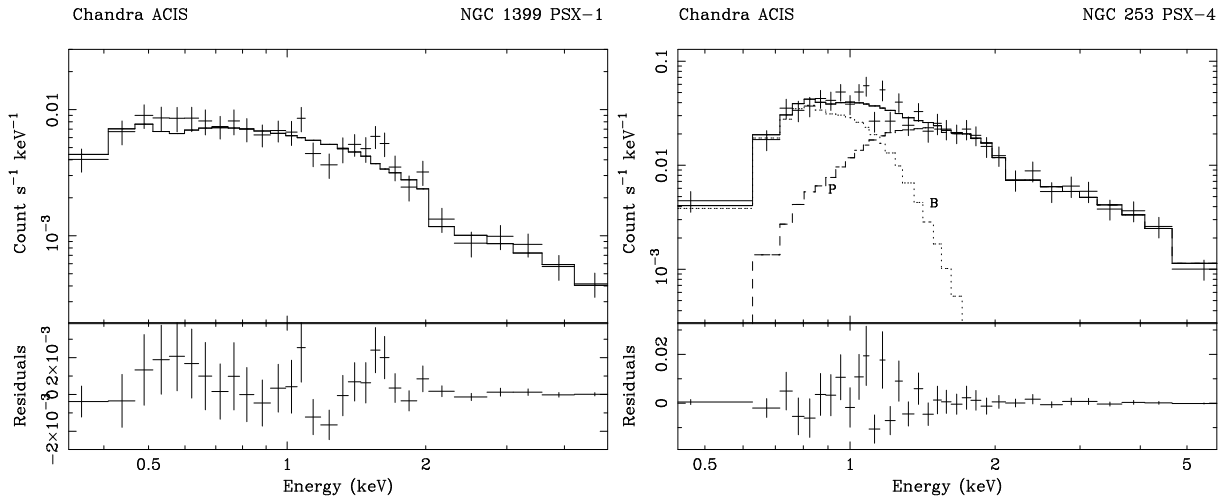
                                     
\includegraphics[width=65mm,height=80mm,angle=270]{f5a}
\includegraphics[width=65mm,height=80mm,angle=270]{f5b}
\caption{Best fits to the spectra of the ULX sources NGC\th 1399
PSX-1 and NGC\th 253 PSX-4, chosen to illustrate higher quality
data from the sample. NGC\th 1399 PSX-1 was fitted by an absorbed
power law model, and NGC\th 253 PSX-4 by an absorbed blackbody
plus power law, labelled B and P.}
\end{figure*}

It is clear that the numbers of ULX and of super-Eddington sources
correlate with 60 $\mu$m flux suggesting a connection with star formation.
Fig. 7 shows that the two elliptical galaxies plotted NGC\th 4636 and NGC\th 4697
apparently follow the same trend as the spiral galaxies, but with generally smaller
number counts per unit area and the smallest infrared surface brightnesses.
The amount of star formation in elliptical galaxies would be expected to be
smaller than in spiral galaxies, assuming that the infrared emission arises
from star formation and not from an active nucleus. A full investigation
of these and other individual sources in our sample will require extensive 
future work; however, we note that NGC\th 4636 is a low-luminosity AGN with
radio jets and is a {\it LINER} (Nagar et al. 2000). NGC\th 4697 is a very 
flattened galaxy which is thought may, in fact, be of S0 type (e.g. Dejonghe et al. 1996).
Of the spiral galaxies in Fig. 7, NGC\th 2681 is a {\it LINER} in a S0 galaxy
with a circumnuclear ring of star formation (Gonz\'alez Delgado et al. 1997).
The galaxy NGC\th 253 with substantial numbers of bright sources is a starburst galaxy
(see Engelbracht et al. 1997). 

\begin{figure*}
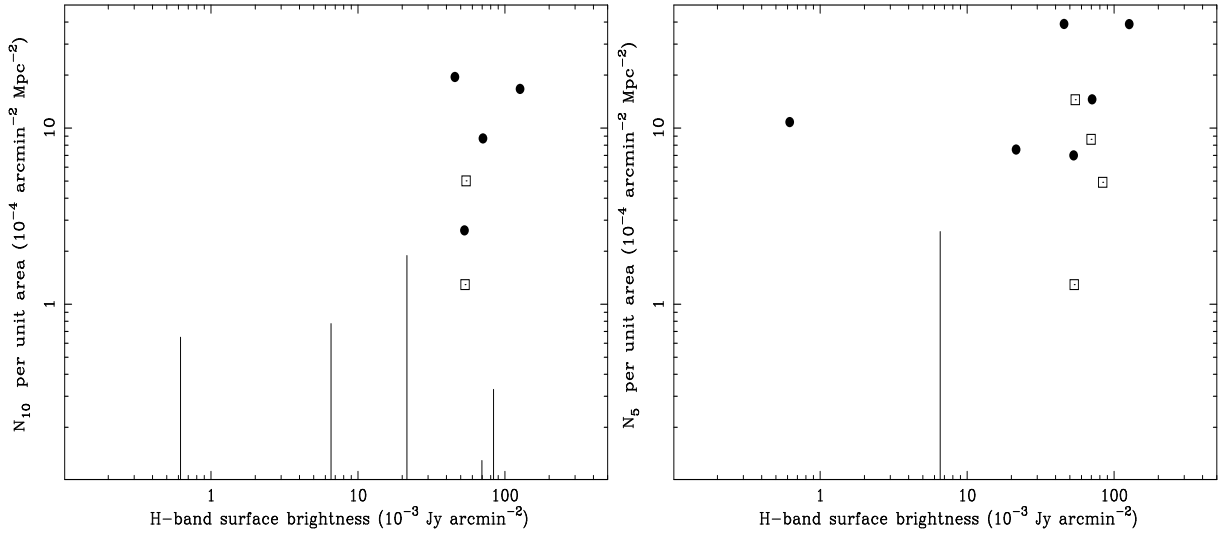
                                        
\begin{center}
\includegraphics[width=70mm,height=80mm,angle=270]{n10_H}  
\includegraphics[width=70mm,height=80mm,angle=270]{n5_H} 
\caption{ULX count per unit projected area (left panel)
and super-Eddington source count per unit projects area
(right panel) {\it versus} {\it H}-band surface brightness.
Open squares show elliptical galaxies and closed circles show
spiral galaxies.}
\end{center}
\end{figure*} 

\begin{figure*}
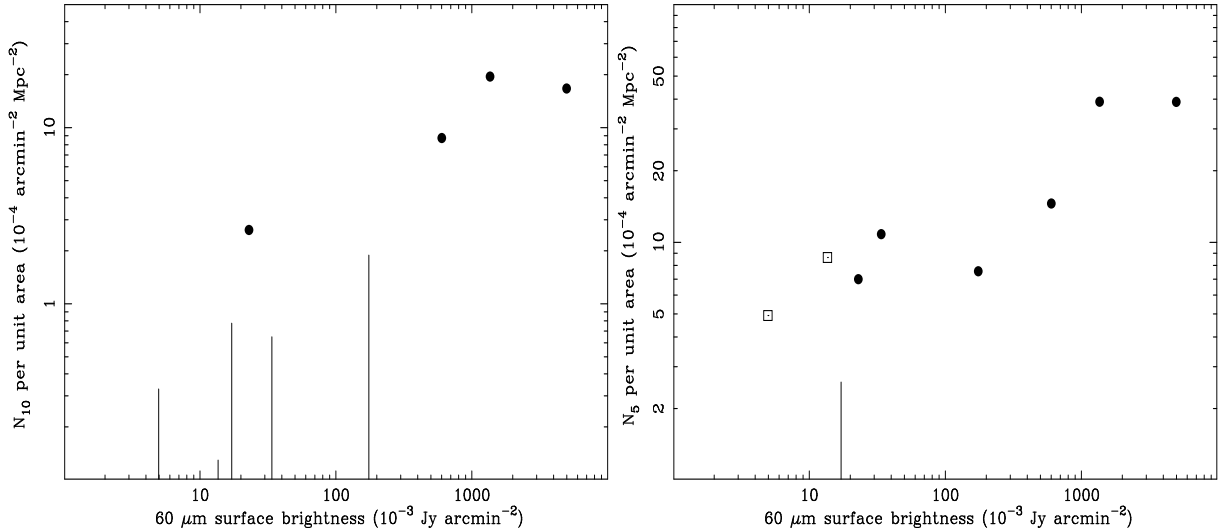
                                        
\begin{center}
\includegraphics[width=70mm,height=80mm,angle=270]{n10_f60}
\includegraphics[width=70mm,height=80mm,angle=270]{n5_f60} 
\caption{$N_{10}$ per unit area (left panel) and $N_5$ per unit area 
(right panel) {\it versus} 
60 $\mu$m surface brightness. Open squares show elliptical galaxies 
and closed circles show spiral galaxies.}

\end{center}
\end{figure*}


\section{Discussion} 

In a survey of 13 normal galaxies we have detected 22 ULX (with
$L_{\rm x}$ $>$ $L_{10}$); of these, 15 exceed $L_{10}$ by less
than a factor of two while the remainder have higher
luminosities, 5 of which are $\approx$ $10^{40}$ erg
s$^{-1}$. Thus the fainter objects are not inconsistent with the
Eddington limit for a 10 M$_{\sun}$ black hole binary, whereas
the brighter objects exceed this limit by factors of 30 -
100. We have found no evidence of correlation with stellar mass
but substantial evidence for a correlation with star forming mass.
Fig. 7 shows that more bright sources are found in spiral
galaxies consistent with the larger star forming mass in these.

\subsection{Nature of the ULX}

In many cases, our results from hardness ratio studies do not
discriminate between types of object. Similarly, because of poor
statistics, spectral fitting does not in many cases reveal the
nature of the object. Consequently, we concentrate discussion on
those cases where the results do help identify the type of
object.

The colour-colour diagrams revealed 2 objects consistent with
High State black hole binary nature: NGC\th 1399 PSX-1 and PSX-2.
Spectral fitting results were obtained for 14 ULX. None of these
required a photon index $>$ 3 suggesting black hole nature;
however, two sources were well-fitted by a one-component
blackbody model (NGC\th 253 PSX-1 and NGC\th 4631 PSX-1), and
this suggests BHB nature. The luminosities of these two sources
were $\rm {1.2\times 10^{39}}$ and $\rm {1.3\times 10^{39}}$ erg
s$^{-1}$, i.e. not exceeding the Eddington limit for a 10
M$_{\sun}$ black hole.  Two-component models similarly did not
provide evidence for black hole nature.

The correlation between ULX and super-Eddington source numbers and 60 $\mu$m
flux supports previous results that such
sources occur preferentially in star forming regions (Zezas et al. 1999;
Roberts \& Warwick 2000; Fabbiano et al. 2001). Moreover, our results
also show detection of sources in elliptical galaxies, and interestingly,
Fig. 7 shows that $N_5$ in the two elliptical galxies NGC\th 4636 and NGC\th 4697
appear to follow the same correlation as for the spiral galaxies,
sugesting that star formation could be important in these two galaxies also.

\subsection{Comparison with previous ULX number counts}

Next, we compare the numbers of detections in the present work
with previous surveys. Roberts \& Warwick (2000) reported
results from the {\it Rosat} HRI for a sample of 486 {\it B}-band
bright northern galaxies, and detected 142 non-nuclear sources.
We have used their luminosity distribution $\rm
 {dN/dL_{38}}$ = $\rm {(1.0 \pm 0.2)\,L_{38}^{-1.8}}$, with $L_{38}$ 
normalized to a {\it B}-band luminosity of 10$^{10}$
L$_{\sun}$ to calculate expected numbers of bright objects in our sample.
In NGC\th 1399 and NGC\th 253 ULX counts of 1.3 and 0.4 are expected,
and 7 and 3$\pm$2 respectively, obtained. In these two galaxies,
$N_5$ values of 2.2 and 0.7 are expected, and 21 and 6.5 obtained.
It appears that there is an underestimation of both ULX and $N_5$ sources in
the {\it Rosat} survey by a factor of 5 -- 10 compared with {\it
Chandra}, indicating the improved sensitivity and angular
resolution of {\it Chandra}.

Grimm et al. (2003) using previous studies of High Mass X-ray Binaries
in the Milky Way and Magellanic Clouds, and {\it Chandra} and {\it ASCA}
observations of nearby starburst galaxies, showed a linear relation
between the star formation rate (SFR) as measured by far infrared flux
and HMXB collective luminosity. They
proposed that the number or collective X-ray luminosity of HMXB could thus
be used as a measure of star formation rate. As an example, we will
compare with the galaxy in our sample having the highest 60 $\mu$m surface 
brightness, the starburst galaxy NGC\th 253 (Fig. 7).
We can adopt a value of $SFR/M_{\rm dyn}$, where $M_{\rm dyn}$ is the
dynamical mass of the galaxy of $\sim 2.0\times 10^{-11}$ yr$^{-1}$
from Grimm et al. for galaxies of similar type (Sc). Taking $M_{\rm dyn}$
as $\rm {1.2\times 10^{11}}$ M$_{\sun}$ (calculated from
the maximum rotation velocity from {\it LEDA}), this gives a value
of $SFR$ $\sim$ 2.4 M$_{\sun}$ yr$^{-1}$ which combined with the
linear relation (Fig. 3 of Grimm et al.) gives a count $N_2$ = 7.
This compares well with our measured value of 9 (Table 3).

In the case of the Antennae galaxies, our results can be used to predict
the expected numbers of bright objects. Using approximate fits to the data in 
Fig. 7 plus the 60 $\mu$m flux of the Antennae galaxies, we estimate 
a value of $N_5$ of 39, and a value of $N_{\rm 10}$ of 21.7
agreeing reasonably with the 29 super-Eddington and 17 ULX
actually detected (Fabbiano et al. 2001).

\subsection{The most luminous sources}

Two X-ray sources have luminosities exceeding 10$^{40}$
erg~s$^{-1}$: NGC\th 1132 PSX-1 and NGC\th 1291 PSX-1. These
luminosities imply a mass of at least 100 M$_{\sun}$ if emitting
isotropically, assuming they are Eddington limited, and it is
important to consider these cases carefully.

The former source has 21 counts, so its luminosity is poorly
determined, and the error-bars are sufficiently large that the
source luminosity may be as small as $\rm {70\times 10^{38}}$ erg
s$^{-1}$.  This is still a very luminous object. It is situated
away from the nucleus of its parent galaxy, and we expect 0.2
background sources of similar apparent brightness (Sect. 3), so
we cannot exclude the possibility that it is not associated with
the galaxy.  NGC\th 1291 PSX-1 has a high count of 390, and a
lower limit luminosity of $\rm 60\times 10^{38}$ erg s$^{-1}$.
It is a point source embedded in extended emission at the centre
of its galaxy, and may be similar to low luminosity AGN-like
cores in many nearby galaxies (Ho et al. 2001).

\subsection{NGC\th 1399 and Globular Cluster X-ray Sources}

NGC\th 1399 is the massive central galaxy in the Fornax Cluster
having an extensive globular cluster system (Kissler-Patig et
al. 1999). Individual globular clusters will not generally be
resolved and so multiple sources within them may be confused.
Indeed Angelini et al. (2001) note a higher average count rate of
apparently point-like sources in globular clusters in NGC\th
1399, which they suggest may be explained in this manner. This
may be the reason why there are so many ULX in this galaxy.
Bright globular cluster X-ray sources are usually LMXB. Although
there is some evidence for globular cluster sources being
somewhat less bright on average than LMXB in the rest of the
Galaxy, Verbunt (2002) concludes that this may be unreal. Thus if
we use an average luminosity of non-cluster sources of $\rm
{2\times 10^{37}}$ erg s$^{-1}$ (Verbunt et al. 1984), we find
that the addition of 50 of these would needed to explain any ULX
in this galaxy.

\subsection{Limits on Non-X-ray Binary Populations}

The super-Eddington sources and the ULX in our sample are
point-like, within the spatial resolution of {\it Chandra},
implying sizes typically $<$ 50 pc. These limits would seem to
preclude super-bubbles or super-giant shells being possible sites
for the emission. Although young SNR may be compact, we would
expect these to be found preferentially in regions of continuing
star formation in late-type galaxies.  SNR spectra may either
appear as optically-thin emission with lines, or alternatively be
featureless.  Our spectral fitting shows Mekal models as
preferred in only one or two cases, suggesting a SNR nature; we
may however detect other SNR with featureless spectra which are
not revealed in this way.  Zezas et al. (2002) show by optical
identification that a relatively small fraction of ULX in their
sample are SNR, implying that our dependence on infrared flux
must be mostly due to other types of object.

\section{Conclusions} 

This survey of 13 galaxies revealed 22 ULX sources with $L_{\rm
x}$(0.3-7.0~keV)$> 1\times 10^{39}$ erg~s$^{-1}$, and a further
39 sources with $L_{\rm x} > 5\times10^{38}$ erg~s$^{1}$ (but
less than $1\times 10^{39}$ erg~s$^{-1}$).  Accidental
coalignment of background sources is not a major effect as 62\%
of these ULX must truly lie within their apparent host galaxies.
Several sources were found that were clearly extended beyond the
instrumental PSF and are shown in Table 5. Two sources appeared
on the colour-colour diagram in a position suggesting possible
black hole nature.

We present spectral fitting results for 14 of the 22 ULX having a
count $>$ 200, the total count varying between 200 and 3200 in
the spectra fitted. Several of the sources could be fitted by an
absorbed power law model suggesting non-thermal emission such as
Comptonization. Seven of the sources were well-fitted by a blackbody
plus power law model, also suggesting X-ray binary nature,
although it was not possible to distinguish the type of XRB. Two
sources were well-fitted by a one-component blackbody model
suggesting High State black hole nature. However, we do not have
positive evidence that any more than a relatively small fraction
of the ULX are black hole binaries.
Moreover, we find evidence that the numbers of ULX and super-Eddington sources
are correlated with the star formation rate as indicated by the 60 $\mu$m flux
supporting similar results found previously, which is consistent with 
a link with younger stellar populations such as high mass X-ray binaries.

\section*{Acknowledgments}
We thank Andreas Zezas and Marek Urbanik for helpful discussions.
PJH thanks the Center for Astrophysics for support.
This work has made use of the CXC archive and the NASA
Astronomical Data Center archive.

\end{document}